\documentclass{aa}

\usepackage{graphicx}
\usepackage{txfonts}
\usepackage{hyperref}
\hypersetup{
    colorlinks=true,
    linkcolor=black,
    urlcolor=black,
    citecolor=blue
    }
\makeatletter
\newcommand*{\rom}[1]{\expandafter\@slowromancap\romannumeral #1@}
\makeatother

\begin{document}

   \title{Origins of coplanar counterrotating stellar disk components in late-type galaxies. \rom{1}}

   \subtitle{}

    \author{M. C. Bugueño\inst{1}
        \and Facundo A. Gómez\inst{1}
        \and Arianna Dolfi\inst{1}
        \and Patricia B. Tissera\inst{2, 3}
    }
    
    \institute{Departamento de Astronomia, Universidad de La Serena, Av. Raúl Bitrán 1305, La Serena, Chile
    \and
    Instituto de Astrofísica, Pontificia Universidad Católica de Chile, Av. Vicuña Mackenna 4860, Santiago, Chile.
    \and
    Centro de Astro-ingenería, Pontificia Universidad Catolica de Chile, Av. Vicuña Mackenna 4860, Santiago, Chile.}
    
    \date{Received 3 October 2025 / Accepted 23 December 2025}
 
  \abstract
   {Understanding galaxy evolution is key to explaining the structures we observe in the present-day Universe. Counterrotating stellar disks, i.e., co-spatial stellar disks rotating with opposite angular momentum, have been proposed as signatures of past accretion events. Therefore, they constitute potential tracers of galactic assembly.}
   {In this work, we aim to investigate the properties, formation channels, and significance of coplanar counterrotating disk components (CRDs) in a sample of Milky Way mass galaxies using the IllustrisTNG cosmological simulations.}
   {We selected an initial sample of 260 central late-type galaxies (i.e. $M_{\rm tot} \approx 10^{12}$, $D/T>0.5$, $N_{\rm star}>10^5$). For each galaxy, we measured the circularity of its stellar particles, and we defined a CRD by considering all particles with circularity $\epsilon < -0.7$, which are located within the spatial extension of the main disk. We then characterized the mass fraction, spatial extent, and star formation history of the CRDs.}
   {Out of the 260 late-type galaxies, we find 26 host significant CRDs (i.e., contributing at least 1\% of the mass of the stellar disk). This means that CRDs are rare, and this outcome is consistent with the results from recent observations. We also find that most of the CRDs are compact (i.e., 88\%) and in situ dominated (i.e., 73\%), and they exhibit bursty star formation histories whose peaks often coincide with external perturbations. This means that external perturbations are able to catalyze retrograde star formation, even when a majority of the CRD's star population is in situ. Finally, we find that a variety of formation pathways can lead to CRDs, including interaction-induced in situ bursts and smooth accretion of misaligned gas.}
   {Overall, our results suggest that CRDs are rare but diverse in origin. In most cases, their formation is linked to the accretion of retrograde gas, either through mergers or environmental inflow, suggesting that these are sensitive tracers of the galaxy's past accretion history.}


   \titlerunning{Origins of CRDs in late-type galaxies. I}
   \subtitle{Identification, characterization, and formation mechanisms of coplanar counterrotating stellar disk components.}
   \maketitle

\section{Introduction}
\label{section:intro}

Present-day galaxies exhibit a wide variety of structural and kinematic properties (e.g., \citealt{2016MNRAS.457.1419M}; \citealt{2018MNRAS.480.2544R}) shaped by both internal  and external processes such as minor mergers, ram-pressure stripping, and galaxy harassment \citep{2006PASP..118..517B}. Each of these mechanisms leaves long-lasting features that can be used to characterize their evolution. A particularly revealing class of features are multi-spin components \citep{1994AJ....108..456R}, that is, kinematic stellar structures whose angular momentum is misaligned with respect to the main stellar disk. These components highlight the complexity of galactic dynamics, as present-day kinematics preserve the imprints of past accretion and interaction events (\citealt{2016MNRAS.457.1419M}; \citealt{2018MNRAS.480.2544R}; \citealt{2019MNRAS.485.2589M}; \citealt{2023MNRAS.526..567D}). Among these misaligned components, co-spatial counterrotating stellar disks \citep{1994ApJ...420..558B} are of particular interest. They consist of two coexisting stellar components that are strongly misaligned in angular momentum but confined to the same plane. Their formation pathways and demographics provide unique constraints on the role of external accretion in shaping galaxy evolution.

The presence of coplanar counterrotating stellar disk components (CRDs) is noteworthy because, if their mass contribution is significant enough, their high velocity amplitudes make it possible to distinguish their stellar populations from those of the main disk, thereby providing insights into the formation and evolution of their host galaxies. It has been argued that counterrotating stellar disks originate from the retrograde acquisition of gas, which settles on misaligned orbits with respect to the preexisting stellar disk and subsequently forms stars (\citealt{2004A&A...424..447P}; \citealt{2010MNRAS.408..783R}; \citealt{2015MNRAS.451.3269V}; \citealt{2019ApJ...878..143S}; \citealt{2021MNRAS.500.3870K}, \citealt{2022ApJ...926L..13B}; \citealt{2024ApJ...973...29B}; \citealt{2025ApJS..281...19G}). The CRDs are therefore direct evidence of past accretion events and constitute valuable tracers of galactic assembly. Several mechanisms have been proposed to explain the origin of the retrograde gas, including accretion from the cosmic web (\citealt{2014MNRAS.437.3596A}), gas-rich mergers with dwarf galaxies (\citealt{1996ApJ...461...55T}; \citealt{2001Ap&SS.276..909P}; \citealt{2014MNRAS.437.3596A}), tidal gas exchange between neighboring galaxies (\citealt{2021ApJS..254...27K}; \citealt{2023AstL...49..229S}), and major mergers that preserve the stellar disk (\citealt{2001Ap&SS.276..909P}; \citealt{2009MNRAS.393.1255C}). Among these formation mechanisms, major mergers appear to be the least common in observed systems.

Observationally, CRDs appear to be rare, with only 1\% of late-type galaxies presenting them (\citealt{2022ApJ...926L..13B}, \citealt{2022MNRAS.511..139B}, \citealt{2025ApJS..281...19G}). At the same time, CRDs are also challenging to observe due to their small mass contribution (\citealt{2021A&A...654A..30R}).
Nevertheless, modern integral field spectroscopy instruments can be used to observe these structures, which are typically identified from the presence of bimodal features in the velocity dispersion maps (\citealt{2022ApJ...926L..13B}; \citealt{2024ApJ...973...29B}; \citealt{2025ApJS..281...19G}). Integral field spectroscopy instruments are also essential to disentangling the two stellar disk components and studying the kinematics and stellar population properties of CRDs in detail, which allows their formation scenarios to be constrained. Complementary techniques, such as spectral energy distribution fitting, are also used to extract physical and chemical properties of CRDs.
Notable examples of CRDs have been identified in early-type disk galaxies. Examples are IC 719 (\citealt{2013ApJ...769..105K}; \citealt{2018A&A...616A..22P}), NGC 4550 (\citealt{1992ApJ...394L...9R}; \citealt{1992ApJ...400L...5R}; \citealt{2013A&A...549A...3C}; \citealt{2013MNRAS.428.1296J}), NGC 3593 (\citealt{1996ApJ...458L..67B}; \citealt{1998A&A...337...80C}; \citealt{2000A&A...363..869G}; \citealt{2013A&A...549A...3C}), NGC 4191 (\citealt{2015A&A...581A..65C}), NGC 448 (\citealt{2016MNRAS.461.2068K}), NGC 1366 (\citealt{2017A&A...600A..76M}), and NGC 5102 (\citealt{2017MNRAS.464.4789M}). However, CRDs have also been observed in late-type disk galaxies, as reported in NGC 7217 \citep{1994ApJ...432..575M}, NGC 4138 (\citealt{1996AAS...189.6804J}; \citealt{2014A&A...570A..79P}), and NGC 5719 (\citealt{2007A&A...463..883V}; \citealt{2011MNRAS.412L.113C}).
These CRDs show a broad range of properties. They are generally observed to have younger stellar populations that corotate with the gas disk (\citealt{2004A&A...424..447P}; \citealt{2014A&A...570A..79P}; \citealt{2022ApJ...926L..13B}; \citealt{2025ApJS..281...19G}), steeper metallicity gradients, and broader metallicity distributions than the main stellar disk (\citealt{2013A&A...549A...3C}; \citealt{2022MNRAS.511..139B}; \citealt{2025ApJS..281...19G}). As previously mentioned, the presence of CRDs is typically linked to past accretion events that induce star formation within the galaxy. Previous works have begun to explore the formation and evolution of CRD components using simulations (\citealt{2021MNRAS.503..726L}; \citealt{2024MNRAS.528.2326S}; \citealt{2025A&A...696A..45P}). For example, using the IllustrisTNG100, \cite{2021MNRAS.500.3870K} analyzed the properties and formation mechanisms of significant CRDs in early-type disk galaxies, finding that stellar counterrotation is mainly a result of an external gas infall. However, it is still unclear what the dominant CRD formation mechanisms are and under what conditions CRDs arise in late-type galaxies, topics that remain underexplored compared to early-type galaxy counterparts, where CRDs are more common (\citealt{1996MNRAS.283..543K}; \citealt{2022ApJ...926L..13B}; \citealt{2022MNRAS.511..139B}; \citealt{2025ApJS..281...19G})

In this work we focus on the search, characterization, and analysis of CRDs in Milky Way (MW)-type galaxies ($M_{\rm tot} \approx 10^{12} M_\odot$) in TNG50 from the IllustrisTNG suite of cosmological hydrodynamical simulations. We aim to determine how common CRDs are in MW-type galaxies as well as the dominant CRD formation mechanisms. In particular, we aim to confirm past formation channels proposed by observations of CRDs. The paper is organized as follows. In Section \ref{section:methods} we give an overview of the TNG50 simulation that we used, define our sample of galaxies, and provide our definition of counterrotation. In Section \ref{section:CRD_identification}, an initial characterization of counterrotating components in the galaxies is shown, and afterward, the selection of significant CRDs is defined. In Section \ref{section:CRDstructuralandassembly}, the structural properties and assembly of the significant CRDs is shown, ending with a classification of the CRDs. We present analysis of the possible formation channels for the CRDs in each classification in Section \ref{section:CRDformation}. We discuss our results in Section \ref{sec:discussion}. Finally, we provide a summary of our results and conclude in Section \ref{sec:conclusions}.

\section{Methodology}
\label{section:methods}

In this section we discuss the simulations employed in this paper. We also describe the galaxy sample selection criteria and the definition of counterrotating stellar disks.

\subsection{The IllustrisTNG: TNG50}
\label{section:illustris}

    \begin{table*}[ht]
        \caption{Parameters of the sample of galaxies with a significant CRD component and the six boundary cases.}
        \label{table:info_CRDs}
        \centering
        \begin{tabular}{lccccccccccc}
            \hline \\
            Name &  SubFindIDat99 &  $M_{\rm Bound}$ &  $M^{\rm CRD}$ &  $M^{\rm Disk}$  &  $R_{50}^{\rm Disk}$ &  $R_{\rm opt}$  & $R_{50}^{\rm CRD}$ &  $R_{95}^{\rm CRD}$ & $R_{95}^{\rm CRD}/R_{50}^{\rm CRD}$ &  Exsitu CRD \\
                         &                &            &        &            &            &           &              &     &      &  fraction       \\
            \hline \\
                       - &              - &  [$10^{10}M_\odot$] &  [$10^{8}M_\odot$] &  [$10^{10}M_\odot$] &  [kpc]     &  [kpc]    &  [kpc]    &  [kpc]   & - &  -              \\
            \hline \\
                    CR-0 &         501208 &     116.20          &               7.76 &                7.52  &  96.93 & 16.50 &      5.36 &     10.83 &  2.02 &             0.11 \\
                    CR-1 &         394621 &     234.03          &               6.00 &                6.00  & 100.03 & 17.00 &      0.52 &      4.05 &  7.82 &             0.03 \\
                    CR-2 &         526478 &      77.48          &               6.73 &                4.13  &  61.40 & 26.50 &      1.61 &      9.06 &  5.63 &             0.33 \\
                    CR-3 &         531910 &      76.65          &               6.56 &                4.97  &  75.68 & 11.00 &      0.83 &      9.72 & 11.75 &             0.17 \\
                    CR-4 &         559036 &      53.31          &               3.02 &                3.16  & 104.50 & 20.50 &      0.69 &      6.61 &  9.52 &             0.24 \\
                    CR-5 &         547844 &      66.70          &               7.52 &                4.78  &  63.55 & 12.00 &      8.00 &     11.26 &  1.41 &             0.12 \\
                    CR-6 &         552879 &      54.49          &               3.15 &                2.96  &  94.15 & 18.50 &      1.46 &      6.70 &  4.58 &             0.62 \\
                    CR-7 &         557396 &      52.82          &               4.02 &                2.71  &  67.41 & 18.00 &      2.38 &      6.95 &  2.92 &             0.37 \\
                    CR-8 &         557721 &      63.53          &               4.75 &                3.92  &  82.43 & 11.00 &      0.51 &      2.55 &  5.00 &             0.01 \\
                    CR-9 &         563732 &      58.77          &               4.79 &                3.49  &  72.85 & 13.00 &      0.54 &      3.65 &  6.81 &             0.02 \\
                   CR-10 &         570319 &      41.79          &               2.55 &                2.62  & 102.68 & 21.50 &      3.52 &     13.86 &  3.94 &             0.13 \\
                   CR-11 &         587019 &      33.28          &               2.10 &                1.89  &  89.89 & 10.00 &      1.05 &      2.94 &  2.81 &             0.01 \\
                   CR-12 &         597311 &      29.82          &               1.17 &                0.81  &  68.77 & 18.00 &      0.98 &      3.31 &  3.39 &             0.01 \\
                   CR-13 &         601861 &      28.97          &               2.25 &                1.63  &  72.57 &  7.50 &      0.80 &      3.68 &  4.60 &             0.00 \\
                   CR-14 &         610988 &      28.61          &               1.28 &                1.23  &  95.66 & 13.50 &      0.68 &      5.06 &  7.46 &             0.18 \\
                   CR-15 &         616825 &      25.00          &               1.17 &                0.67  &  57.57 & 12.00 &      0.94 &      4.26 &  4.52 &             0.26 \\
                   CR-16 &         617324 &      23.96          &               1.13 &                0.98  &  86.86 & 12.00 &      1.61 &      4.95 &  3.07 &             0.03 \\
                   CR-17 &         619801 &      22.74          &               3.15 &                0.74  &  23.48 &  8.00 &      2.20 &      5.20 &  2.36 &             0.01 \\
                   CR-18 &         626864 &      21.85          &               2.07 &                0.94  &  45.52 & 14.50 &      1.52 &      6.89 &  4.54 &             0.01 \\
                   CR-19 &         627459 &      22.74          &               1.43 &                1.24  &  86.97 & 13.00 &      0.89 &      4.72 &  5.31 &             0.37 \\
                   CR-20 &         627572 &      23.44          &               4.34 &                1.14  &  26.25 & 10.50 &      0.79 &      3.08 &  3.90 &             0.02 \\
                   CR-21 &         628031 &      24.00          &               2.98 &                0.87  &  29.07 &  7.00 &      0.90 &      3.32 &  3.70 &             0.01 \\
                   CR-22 &         594887 &      33.48          &               1.56 &                1.62  & 103.79 & 18.00 &      2.84 &      5.41 &  1.90 &             0.03 \\
                   CR-23 &         603281 &      22.02          &               0.62 &                0.60  &  96.17 & 14.00 &      0.78 &      5.13 &  6.55 &             0.22 \\
                   CR-24 &         617559 &      25.05          &               0.95 &                0.95  & 100.08 & 13.00 &      2.05 &      5.93 &  2.89 &             0.03 \\
                   CR-25 &         623575 &      22.01          &               1.06 &                1.10  & 104.09 & 12.00 &      1.83 &      5.73 &  3.14 &             0.09 \\
            \hline
        \end{tabular}
        \tablefoot{
        Column descriptions from the left to the right are: (1) internal name, (2) z=0 SubfindID from the TNG50 simulations, (3) total gravitationally bounded mass as defined by the Subfind algorithm, (4) stellar mass of the CRD, (5) total stellar disk mass, (6) half-mass radius of the stellar disk, (7) optical radius, (8) half-mass radius of the CRD, (9) radius enclosing 95\% of the CRD, (10) stellar concentration of the CRD as defined in Sec. \ref{section:CRDstructuralandassembly}, (11) fraction of ex situ stellar particles in the CRD.
        }
        \end{table*}

    The IllustrisTNG project \citep{2015A&C....13...12N} is a set of gravitational magnetohydrodynamics cosmological simulations ran with the moving-mesh code AREPO \citep{2010MNRAS.401..791S}. The project is made up of three simulation volumes: TNG50 (\citealt{2019MNRAS.490.3234N}, \citealt{2019MNRAS.490.3196P}), TNG100, and TNG300 (\citealt{2018MNRAS.475..648P}, \citealt{2018MNRAS.475..676S}, \citealt{2018MNRAS.475..624N}, \citealt{2018MNRAS.477.1206N}, \citealt{2018MNRAS.480.5113M}). Each of these runs solves the coupled evolution of dark matter, gas, stars, and supermassive black holes from $z=127$ to the present day, $z=0$. All of the simulations of the IllustrisTNG project are governed by a Lambda cold dark matter model with a cosmology consistent with the results of \cite{2016A&A...594A..13P}. The IllustrisTNG set of simulations has produced results consistent with observations in regimes beyond the ones used to calibrate the model \citep{2015A&C....13...12N}. For this work, we used the TNG50 run, and more specifically, we analyzed the baryonic simulation, TNG50-1 (TNG50, hereafter), which is the highest resolution simulation of the IllustrisTNG project. Its physical simulation box size is roughly 50 Mpc per side in length, and the stellar and dark matter particle mass resolutions are $8.5 \times 10^{4}$ M$_{\odot}$ and $4.5 \times 10^{5}$ M$_{\odot}$, respectively. The softening lengths at $z=0$, are $\epsilon_{{\rm DM},*} = 288$ pc and   $\epsilon_{gas,min} = 74$ pc. This simulation allows the components of galaxies to be analyzed in more detail than the other available simulations.

    The TNG50 data provide the positions, velocities, mass, metallicity, and metal abundances of the stellar particles as part of the complete snapshot data. Each stellar particle represents a single stellar population with a given age and metallicity that is hereditary from the progenitor gas cell. Information about the galaxy and time where the stellar particle is born is provided in \cite{2015MNRAS.449...49R} and is described in more detail in Section \ref{section:crdef}. The TNG50 simulation also gives information about the galaxies extracted from their member particles as a whole, classified according to a friends-of-friends (FoF) algorithm, and subsequently Subfind, including their total mass, total stellar mass, and virial radius. Information about the disk mass fraction according to the circularity of the particles is provided as an additional catalog by \cite{2015ApJ...804L..40G}.

\subsection{Sample selection}
\label{section:sample}
    In this study, we focus on a sample of MW-type galaxies; in a follow up study, we will expand this sample to include a wider range of late-type galaxies.
    In particular, we selected galaxies that fulfill the following criteria:
    \begin{itemize}
        \item A total mass range of $10^{11.5} - 10^{12.5} M_{\odot}$
        \item A disk-to-total mass ratio (D/T) larger than $0.5$.
        \item A total number of stars $N_{\rm tot, stars} \ge 10^{5}$
        \item The galaxy must be identified as the central galaxy of its FoF halo, defined as the most massive subhalo within that FoF group.
        
    \end{itemize}
    In particular, the D/T ratios were obtained from the publicly available database of IllustrisTNG as part of the additional catalogs\footnote{\url{www.tng-project.org/data/docs/specifications/\#sec5c}} and were computed by considering as disk stellar particles the ones with a circularity parameter (see Sec \ref{section:crdef})  larger than 0.7.
    
\begin{figure}[t]
        \centering
        \includegraphics[width=\columnwidth]{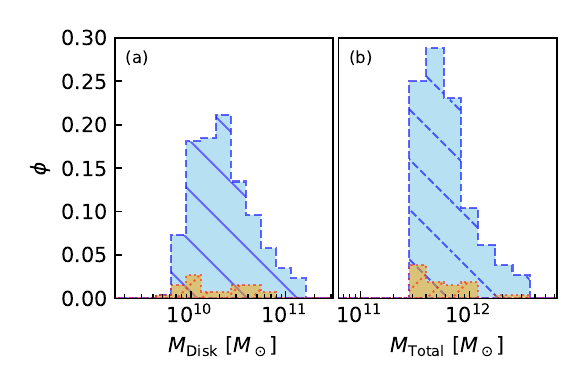}
        \caption{Distributions of mass of the full sample of galaxies and the galaxies hosting a significant CRD. Panel (a): Distribution of stellar mass confined within the disk region, independent of the circularity parameter. Panel (b): Distribution of the total mass according to the Subfind algorithm (i.e., considering all particles bound to the galaxy). In both panels, the distribution was normalized according to the full sample of 260 galaxies, represented by the light blue shaded distribution. The orange shaded region represents the distribution of the 26 galaxies hosting a significant CRD component.}
        \label{fig:diskandtotal_mass_distribution}
    \end{figure}

    After applying the described selection criteria, we obtained a total of 260 late-type disk galaxies. As discussed, the selected galaxies have $\geq 5 \times 10^{5}$ stellar particles within them, thus ensuring enough particles are present to reasonably resolve the structure of the stellar disk and search for potential structures with small mass contributions, such as counterrotating stellar components. The light blue histograms in Figure \ref{fig:diskandtotal_mass_distribution} show the distribution of stellar mass within the galactic disk (left panel, as defined in Sec. \ref{section:crdef}) and of the total mass (right panel) of the full sample of the full sample of 260 initially selected galaxies.

    \begin{figure}[t]
        \centering
        \includegraphics[width=\columnwidth]{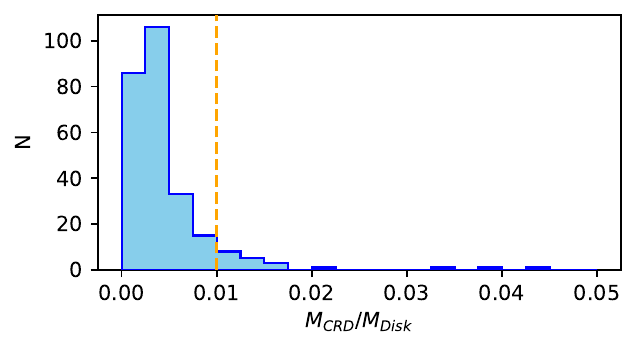}
        \caption{Distribution of the coplanar counterrotating disk component mass fraction for the full sample of 260 late-type disk galaxies. The vertical dashed orange line indicates a 1\% mass fraction from the CRD. Only 10\% of the galaxies (when including the boundary cases) have a CRD mass fraction greater than 1\%.}
        \label{fig:crfraction_distribution}
    \end{figure}

    \begin{figure*}[hbt]
        \centering
        \includegraphics[width=6.8in]{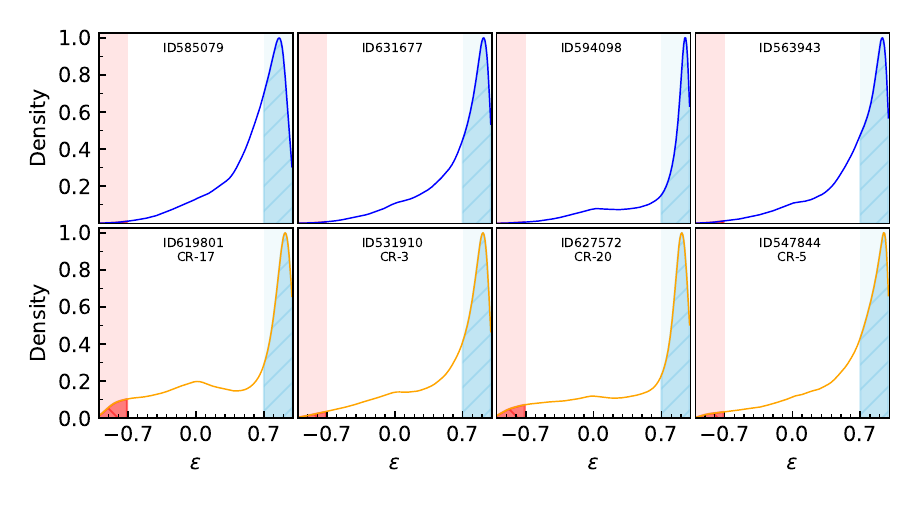}
        \caption{Distribution of the circularity parameter for eight galaxies from the selected MW-type sample. The curves have been normalized so that the peak value is equivalent to one. The four top panels are galaxies that do not possess a significant CRD, while the bottom panels come from the subsample of 26 galaxies with a significant CRD. The shaded regions represent star particles identified as the CRD component (red) and the main stellar disk (light blue).}
        \label{fig:circularity_examples}
    \end{figure*}

\subsection{Coplanar counterrotating disk component definition}
\label{section:crdef}

    The goal of this work is to identify, quantify, and analyze CRDs in fully cosmological simulations of galaxy formation. Different methods for isolating particles associated with different galaxy components, such as counterrotating disks, have been proposed in previous works. For example, \cite{2012MNRAS.420..255T} applied a criteria based on circularity and binding energy to subdivide a late-type galaxy based on its different components. In this study, we follow an approach more closely related to the definitions used in \cite{2017MNRAS.472.3722G} and \cite{2019MNRAS.485.2589M}. We first rotated the galaxy so that the plane of the disk is coplanar with the xy-plane, i.e., the angular momentum of the galactic disk is parallel to the z-axis direction. Afterward, we identified particles that are counter-rotating in the galaxy using the circularity parameter, thus performing a kinematic decomposition. Following \cite{2003ApJ...597...21A}, this parameter is defined as

    \begin{equation}
    \epsilon_{J_z} = \frac{J_z}{|\mathrm{max}(J_z(E))|},
    \label{eq:circularity}
    \end{equation}

    where $J_z$ is the angular momentum in the z-axis and the term "$\mathrm{max}(J_z(E))$" indicates the maximum angular momentum in the z-direction allowed for that orbital energy, i.e., that of a circular orbit.

    To define the extension of the stellar disks, we used the optical radius, $R_{\rm opt}$. As in \citealt{2023MNRAS.523.5853V} and \citealt{2023MNRAS.526..567D}, $R_{\rm opt}$ is defined as the radius at which the surface brightness profile on the V-band of the galaxy, orientated in a face-on projection, reaches $\mu_V=25\ \mathrm{mag\ arcsec}^{-2}$. To estimate the optical radius, we measured the surface brightness of the galaxy in cylindrical shells of width 0.5kpc and height 10kpc above and below the disk plane. The values of the $R_{\rm opt}$ obtained for each galaxy are listed in Table \ref{table:info_CRDs}.

    \begin{figure*}[t]
        \centering
        \includegraphics[width=7in]{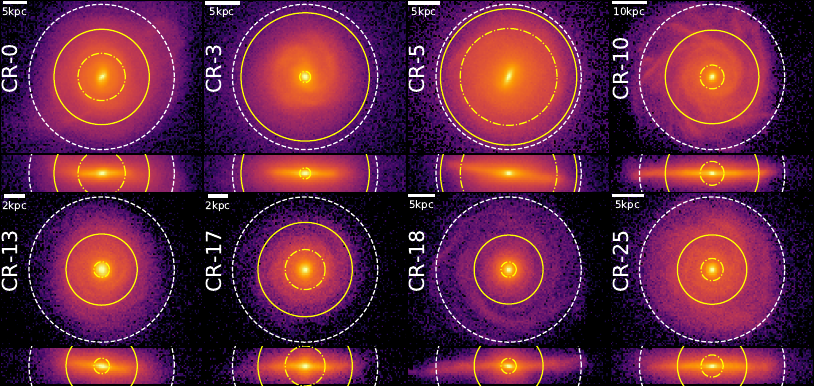}
        \caption{Face-on and edge-on projections of the stellar mass surface density for a subsample of eight galaxies with the most significant counterrotating stellar disks. The galaxies are sorted according to their total mass, from the most to the least massive. We show all the stellar particles that belong to the host galaxy. The solid and dot-dashed yellow circles represent the radii, $R_{95}^{\rm CRD}$ and $R_{50}^{\rm CRD}$, of the CRD, respectively, while the dashed white circle represents the optical radius, $R_{\rm opt}$, of the galaxy.}
        \label{fig:8subhaloes_all}
    \end{figure*}

    \begin{figure*}[thb]
        \centering
        \includegraphics[width=7in]{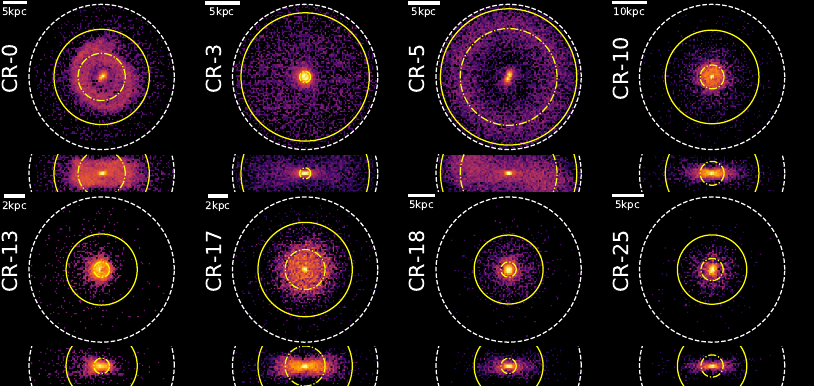}
        \caption{Same as Fig. \ref{fig:8subhaloes_all}, but we show only the stellar particles belonging to the CRDs.}
        \label{fig:8subhaloes_cr}
    \end{figure*}

    Having the circularity parameter computed for every stellar particle within $R_{\rm opt}$, we then defined the coplanar CRDs by selecting stellar particles that fulfill the following criteria:
    \begin{itemize}
        \item Circularity value in the range of $\epsilon < -0.7$.
        \item A distance from the galactic plane of less than 5kpc ($|z| < 5 \ \rm kpc$).
        \item A galactocentric distance less than $R_{\rm opt}$.
    \end{itemize}
    These selection criteria allowed us to select star particles in nearly circular orbits rotating in the opposite direction of the overall disk. This definition is stricter than what has been used in past works, such as \cite{2021MNRAS.500.3870K}, and is closer to that of \cite{2022MNRAS.509.1764J} and \cite{2021MNRAS.503..726L} (more details in Appendix \ref{sec:appendix_differences_with_past_studies}).

    In this work we are also interested on characterizing the formation mechanisms of CRDs. Due to this, it is important to identify the origins of the different stellar particles associated to these components. In particular, we subdivide the counterrotating stellar particles as in situ or ex situ based on where the particle were born. As discussed in \citealt{2016MNRAS.458.2371R}, a particle is considered in situ if the galaxy in which it formed is part of the main branch of the merger tree of the galaxy where the particle is at the present-day. In case the particle was born outside this main branch then the particle is tagged as ex situ. Notice that particles that are formed within the potential well of the main progenitor from recently stripped gas from satellite galaxies are labeled as in situ.

\section{Identification of coplanar counterrotating disk components}
\label{section:CRD_identification}
In this section we quantify and characterize the CRD components in the sample of 260 late-type galaxies. We then quantify the number of galaxies hosting a significant CRD. Once identified, CRD components are characterized based on three parameters:
\begin{itemize}
    \item The counter-rotating disk mass fraction, $M_{\rm CRD}/M_{\rm Disk}$, which is given as the fraction of mass in the counterrotating component over the total disk stellar mass enclosed within $R_{\rm opt}$ and $|z| \le 5 \ \rm kpc$
    \item $R_{50}^{\rm CRD}$, which quantifies the half-mass radius of the distribution of counterrotating stellar particles.
    \item $R_{95}^{\rm CRD}$, which quantifies the radius at which the mass distribution of counterrotating stellar particles reaches 95\% of the mass of the CRD.
\end{itemize}
Fig. \ref{fig:crfraction_distribution} shows the distribution of the CRD mass fraction obtained from our full sample of late-type disk galaxies. Note that for the majority of disk galaxies, the CRD components do not surpass the $1\%$ threshold, which is highlighted by the vertical dashed line in the figure. We find that only $\approx8\%$ of the galaxies host a CRD with a mass fraction larger than $1\%$. The sample of galaxies that host a significant CRD, i.e., $M_{\rm CRD}/M_{\rm Disk} \ge 0.01$ consists of 20 galaxies. In this study, we decided to include in our final sample six boundary cases with $0.0095 \le M_{\rm CRD}/M_{\rm Disk} < 0.01$. The mass distribution of the final sample is shown in Figure \ref{fig:diskandtotal_mass_distribution} with orange histograms.

In Table \ref{table:info_CRDs} we list the internal properties of the galaxy sample with identified CRDs. In the left-most column, the label given to each identified CRD is shown. The corresponding Subfind ID at $z=0$ is given in the second column. In Figure \ref{fig:circularity_examples} the circularity parameter distribution for eight galaxies is shown. The top four examples in the figure show galaxies that do not host a CRD, while the bottom four examples show galaxies hosting a CRD, according to our definition. The red and light blue shaded regions represent the star particles belonging to the CRD and stellar disk, respectively.

Figure \ref{fig:8subhaloes_all} presents eight examples of galaxies hosting a significant CRD. The figure shows their total stellar surface mass density both in face-on and edge-on projections. We show the most extensive CRDs in terms of their $R_{95}^{\rm CRD}$. We note that galactic disks present a morphological diversity, with some exhibiting well-defined bars (e.g., CR-5) and spiral structures (e.g., CR-10 and CR-18).

Figure \ref{fig:8subhaloes_cr} shows the spatial distribution of the counterrotating stellar particles considered as members of the CRD in the same order as in Fig. \ref{fig:8subhaloes_all}. By comparing the two figures, we observed that in some galaxies (e.g., CR-5 and CR-3), the CRDs extend to radial distances comparable to those of the overall stellar disk. In contrast, other cases (e.g., CR-25) exhibit much more compact CRDs. In addition, some cases such as CR-0 and CR-5 display an interesting torus-like morphology. We explore the origin of their structure in the following sections. The full set of 26 galaxies analyzed is shown in Fig. \ref{fig:CRDs_all_appendix} and \ref{fig:CRDs_CRD_appendix}.

From this analysis, and based on our criteria, it follows that approximately 10\% (including the boundary cases) of the galaxies host a significant stellar CRD. Our results indicate that CRDs are relatively infrequent. This is consistent with previous observational studies, which report CRDs, specially stellar, as rare systems found in only $\approx1\%$ of galaxies found in Mapping Nearby Galaxies at APO (MaNGA) (\citealt{2022ApJ...926L..13B}, \citealt{2025ApJS..281...19G}).

\section{Characterization of CRDs}
\label{section:CRDstructuralandassembly}
In this section we characterize and quantify the main properties of the stellar CRDs. We focus on their radial distributions and the origin of their stellar particles. 

\begin{figure}[hbt]
        \centering
        \includegraphics[width=\columnwidth]{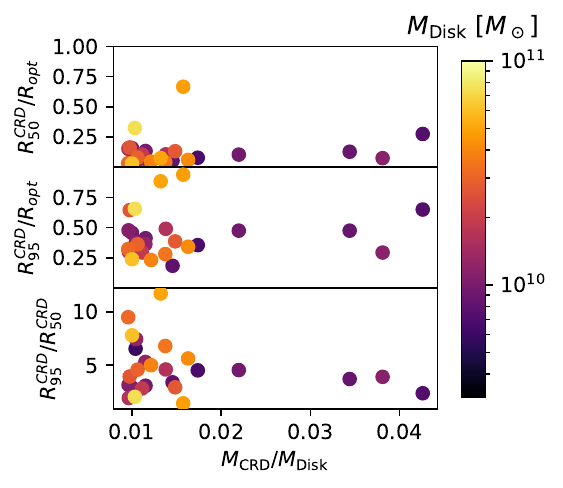}
        \caption{Coplanar counterrotating disk (CRD) mass fraction, $M_{\rm CRD}/M_{\rm Disk}$, as a function of different measurements of the counterrotating stellar disk extension and mass distribution for the sample of 26 significant CRD galaxies. Top and middle panels: Radius containing 50\% and 95\% of the counterrotating stellar disk mass normalized by the optical radius of the galaxy, respectively. Bottom panel: Stellar concentration of the coplanar counterrotating disk component. All data points are color coded by the total stellar disk mass. We observed that while most CRDs are compact, with the majority having $R_{50}^{\rm CRD}/R_{\rm opt} < 0.25$, there are a few extended cases. Overall, we find that our sample of counter rotating galaxies show a variety of structural properties.}
        \label{fig:significantCRDsin3panels}
    \end{figure}

Figure \ref{fig:significantCRDsin3panels} shows the relationship between the CRD mass fraction, $M_{\rm CRD}/M_{\rm Disk}$, and different measurements of the disk extension and mass distribution. In particular, we focus on $R_{50}^{\rm CRD}/R_{\rm opt}$ (top), $R_{95}^{\rm CRD}/R_{\rm opt}$ (middle), and $R_{95}^{\rm CRD}/R_{50}^{\rm CRD}$ (bottom). Data points are color coded according to the total stellar mass of the disk.

The top panel of Fig. \ref{fig:significantCRDsin3panels} shows that most CRDs are relatively compact in nature, with approximately 88\% having a half-mass radius ($R_{50}^{\rm CRD}$) smaller than a quarter of their optical radius ($R_{\rm opt}$). However, the middle panel of the figure reveals that around 89\% of CRDs have $R_{95}^{\rm CRD}/R_{\rm opt}$ values greater than 0.25, and approximately 19\% exceed half the optical radii. This indicates that while most CRDs are compact in terms of their half-mass radii, many have a reasonably extended radial distribution. The bottom panel of Fig. \ref{fig:significantCRDsin3panels} shows  a measure of the CRDs concentration, which is defined as the ratio of $R_{95}^{\rm CRD}/R_{50}^{\rm CRD}$. The CRDs with $R_{95}^{\rm CRD}/R_{50}^{\rm CRD} \leq 2$ contain an approximate significant portion of their mass in a thin ring-like structure. We identified three of these torus-like CRDs, and they are CR-0, CR-5, and  CR-22. Indeed, this can be appreciated on Fig. \ref{fig:8subhaloes_cr}. The CRDs with higher concentration values, such as CR-3 and CR-10,  exhibit less compact profiles. The wide spread in the CRD concentration highlights the structural diversity within the  sample. Overall, we find that our sample of counter rotating galaxies shows a variety of structural properties. Three CRDs show a ring-like structure, some display an extended CRD, and others have a more compact nature with values of $R_{95}^{\rm CRD}$ below half of the disk's $R_{\rm opt}$.

\begin{figure*}[hbt]
        \centering
        \includegraphics[width=7.2in]{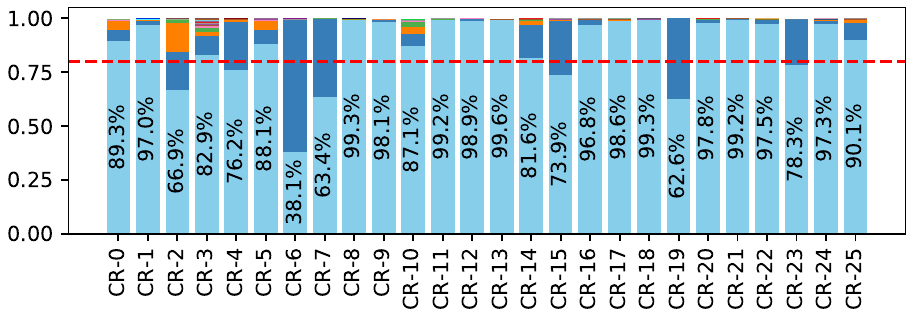}
        \caption{Fraction of the in situ and ex situ stellar particles in the coplanar counterrotating disk component for the sample of 26 significant CRD galaxies. Light blue bars represent the fraction of in situ stellar particles in the CRD, while the remaining colors represent the contribution to the ex situ fraction in the CRD from different satellite galaxies. Each color refers to a different satellite. The dashed red line indicates an in situ contribution equal to 80\%.}
        \label{fig:assemblyinfo_cr}
    \end{figure*}

Next we characterize the origin of the stellar particles that conform each CRD. We recall that stellar particles formed from the gas distribution bound to the potential well of the main host galaxy are defined as in situ. This includes gas that may be recently stripped from closely interacting satellite galaxies. In Fig. \ref{fig:assemblyinfo_cr}, we show the fractions of in situ and ex situ stellar particles in each significant CRD. The light blue (bottom) fraction of each bar indicates the contribution from in situ stellar particles, while the other colored fractions indicate the contribution to the ex situ stellar particles from different satellite galaxies. Interestingly, a majority of the CRDs present a dominant in situ contribution, with 19 out of 26 (73\%) cases displaying fractions above 80\%. The remaining seven CRDs present a more significant ex situ contribution, i.e., $> 20\%$.  We found only one CRD dominated by the ex situ component (CR-6). In general, CRDs with significant ex situ contributions have only one significant progenitor that contributes over 90\% of the ex situ particles, with  CR-2 being the only exception. We explore these cases further in Section \ref{section:CRDformation}.

In general, we find that CRDs have a very significant in situ contribution and, in most cases, present a compact distribution. We note, however, that some show extended configurations. As we show in Section \ref{section:CRDformation}, even though most CRD stellar particles are born within the host potential well, the corresponding star-formation bursts are typically associated with significant interaction with massive nearby satellites.

\begin{figure}[hbt]
        \centering
        \includegraphics[width=\columnwidth]{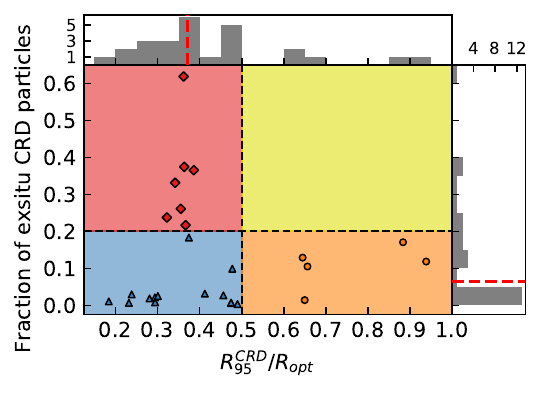}
        \caption{Classification of the sample of 26 significant CRDs according to their counterrotating stellar disk extension ($R_{95}^{\rm CRD}/R_{\rm opt}$) and to their ex situ fraction of stellar particles. We classified the CRDs into four different types, namely, extended ex situ (yellow quadrant), compact ex situ (red rombus quadrant), compact in situ (blue triangle quadrant) and extended in situ (orange circle quadrant).}
        \label{fig:classificationtypes}
    \end{figure}

In Fig. \ref{fig:classificationtypes} we show the distribution of CRDs in a space defined by the fraction of ex situ material and the extension of the CRD with respect to the overall stellar disk, $R_{95}^{\rm CRD}/R_{\rm opt}$. Each symbol represents a single CRD. The top histogram in the figure highlights that a significant fraction of CRDs ($\approx 80\%$) fall into a compact configuration, with values of $R_{95}^{\rm CRD}/R_{\rm opt} < 0.5$. Indeed, the median value of the distribution is $R_{95}^{\rm CRD}/R_{\rm opt} \sim 0.37$. However, we note the presence of the five interesting cases with very extended in situ disks, i.e., $R_{95}^{\rm CRD}/R_{\rm opt} > 0.6$  (see also Fig. \ref{fig:significantCRDsin3panels}). As previously discussed, the histogram in the right panel of the figure shows that most CRDs ($\approx 73\%$) have an ex situ fraction of less than $0.2$. Indeed, the median value of the ex situ mass fraction is approximately $0.06$.

Based on the space represented by $R_{95}^{\rm CRD}/R_{\rm opt}$ and the ex situ fraction, we classified the CRDs into different categories using a threshold of $R_{95}^{\rm CRD}/R_{\rm opt} = 0.5$ and an ex situ fraction of 20\%. The different categories are defined as (i) compact in situ, (ii) compact ex situ, and (iii) extended in situ. These categories reflect both their spatial extent and whether the ex situ mass fraction is significant. The classification scheme is highlighted with different colored regions in Fig. \ref{fig:classificationtypes}. Using this scheme, we found that compact in situ CRDs are the most common, with 14 cases, followed by seven compact ex situ and five extended in situ CRDs. Note that no CRDs with a significant ex situ fraction were found to be extended.

This analysis indicates that CRDs are not only rare in TNG50, but they also predominantly exhibit compact spatial distributions. Interestingly, despite being less frequent, it is the extended types that are more likely to be detected observationally, providing context for the $\approx 1\%$ occurrence reported in the MaNGA survey (\citealt{2022ApJ...926L..13B}, \citealt{2025ApJS..281...19G}). 
Using this CRD classification as a foundation, in the next section we proceed to analyze each type individually, examining their star formation histories (SFHs) and merger activity in order to characterize the nature of their formation.

\section{Formation channels for CRDs}
\label{section:CRDformation}

In this section we explore the different formation channels of the 26 significant CRDs, subdivided according to the three categories defined in Section 4. In our analysis we incorporate information about the stellar ages as well as the properties of infalling satellites.

Figure \ref{fig:ages_CRDs}  shows  the SFHs of all  26 significant CRDs. Each panel represents a galaxy from our sample. The orange and the light blue lines in the figure represent the SFH of the CR ($\epsilon < -0.7$) and the corotating ($\epsilon >  0.7$) disk component, respectively. The vertical dashed lines indicate the median of each distribution. Additionally, the dashed gray line represents the SFH of all stellar particles confined within a cylinder of radius $R = R_{\rm opt}$ and height $|z| \leq 5$ kpc. All SFHs have been normalized to the peak value of the corresponding distribution. It can be clearly seen that with respect to the full disks, CRDs typically have a more bursty history. In most cases, these bursts are also noticeable in the overall disk's SFH. Additionally, we also noticed that CRDs are significantly older than the corotating counterparts. Indeed, this is the case for $\approx93\%$ of the identified CRDs. The two exceptions are CR-0 and CR-11. We study these cases further later in this section.

Figure \ref{fig:ages_CRDs} also highlights the more significant interaction events each galaxy has undergone. We focus on interactions associated with satellites that have a total peak mass of at least $5 \times 10^{9}$ M$_\odot$ and have a satellite to host mass ratio larger than $20\%$ at the corresponding peak mass time. In the figure, the first pericenter passage of such interactions is indicated with arrows, color coded according to the peak mass of each satellite. For CRDs with an ex situ fraction of $\gtrsim 10\%$, we highlight the arrow associated with the most significant ex situ contributor. As expected, in all cases we observed a higher frequency of significant mergers at early times. Nonetheless, significant interactions can take place at any time. Such interactions are commonly associated with star-formation bursts in the overall stellar disk (see also \citealt{2017MNRAS.472.4133G}).

\subsection{Compact in situ CRDs}
\label{sec:compactin situCRD}

\begin{figure*}[hbt]
        \centering
        \includegraphics[width=7.2in]{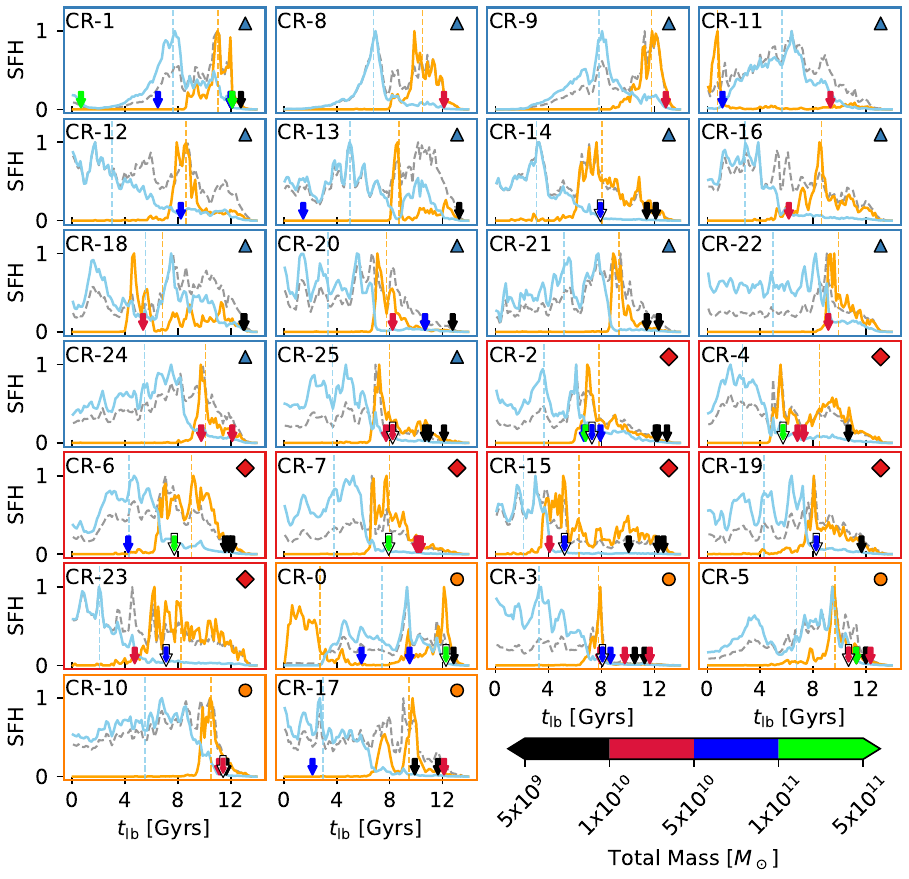}
        \caption{Star formation histories of all 26 significant CRDs, grouped according to the classification type, namely, compact in situ, compact ex situ, and extended in situ (see Sec.\ref{section:CRDstructuralandassembly}). The orange and the light blue lines represent the SFH of the counterrotating and the corotating disk component, respectively. The vertical dashed lines indicate the median of each distribution. The gray line represents the SFH of all stellar particles confined within the disk region independent of their circularity parameter. All SFHs were normalized to the peak value of the corresponding distribution. For each CRD, we also label the time of the first pericenter for the most significant interactions experienced by the galaxy, considering only satellites with a peak total mass greater than $5\times10^{9} M_\odot$ and a satellite-to-host mass-ratio greater than 20\%. The most significant interaction of the CRD is identified by an arrow outlined in black and white.}
        \label{fig:ages_CRDs}
    \end{figure*}

Compact CRDs are the most frequent type in our sample. By definition, these systems are not very extended (their values of $R^{\rm CRD}_{95}$ lying under half of the optical radius of the overall disk), meaning the counterrotating stars are largely confined to the spatial limits of the galaxy’s bulge or bar. Within this category, we first considered the CRDs that formed predominantly in situ (ex situ fraction < 20\%). There are 14 such cases in our sample. In Fig. \ref{fig:ages_CRDs}, the compact in situ CRDs are indicated by blue boxes as well as upward-pointing blue triangle symbols in the top-right corner of each panel.

Among the compact in situ CRDs, one galaxy, CR-14, stands out as a borderline case. About 18\% of its CRD stellar particles are ex situ (accreted). CR-14 lies just below our threshold for the ex situ category, and its counterrotating component can mainly be associated with a significant accretion event. Indeed, Fig. \ref{fig:ages_CRDs} shows that a merger with a satellite with a peak mass on the order of $\approx8\times10^{10}$  M${_{\odot}}$ induces a star-formation burst that gives rise to most of the corotating component. Furthermore, the assembly information shown in Fig. \ref{fig:assemblyinfo_cr} for the CRD of CR-14 shows that the majority of its few accreted stars came from this single infalling satellite (see Fig. \ref{fig:ages_CRDs}, blue highlighted arrow), which contributed stars and gas and induced a burst of star formation.

Many of the compact in situ CRDs show clear signs of interaction-triggered star formation in their histories, even though the stars themselves formed in situ. For example, CR-11, CR-12, CR-14, CR-18, CR-20, and CR-25 all exhibit enhanced star formation in their CRDs that was preceded by an interaction event either at intermediate (e.g., CR-12 and CR-20) or at late times (CR-11). In these galaxies, a passing or merging satellite  disturbed the gas in the host, inducing a burst of star formation in the inner galactic region that contributed stellar particles on retrograde orbits. In each of these cases, the CRD’s stellar mass is almost entirely in situ, and the interactions served as a catalyst for star formation rather than directly depositing a large number of stars.

On the other hand, some compact in situ CRDs appear to have formed very early, alongside the initial assembly of the galaxy’s bulge, and without any later significant interactions. Galaxies such as CR-1, CR-8, CR-9, CR-21, CR-22, and CR-24 exhibit strong early bursts of star formation that are occasionally correlated with the high merger rates characteristic of that epoch. During this early phase of galaxy assembly, star formation occurred in a more irregular and turbulent environment, resulting in a pressure-supported stellar population. The more isotropic kinematics of these stellar particles are associated with broad circularity distributions, contributing to both the corotating and counterrotating components of the final stellar disk. Accordingly, the stellar ages of these CRDs trace back to the formation of the primordial inner regions of the galaxies. These structures are therefore old and centrally concentrated.

The two remaining cases have peculiar formation mechanisms. CR-16  formed over a prolonged period during which the corotating component experienced minimal star formation. Until the merger of an infalling satellite, the galaxy lacked a dynamically cold stellar disk. The CRD is thus a remnant of this earlier evolutionary stage when the galaxy was pressure-supported rather than rotation-supported. Finally, CR-13  formed within a short time window following the smooth accretion of misaligned cold gas. The accreted gas triggered a burst of star formation in the central region, giving rise to the CRD.

In summary, the compact in situ CRDs encompass a variety of formation paths. Some (CR-1, CR-8, CR-9, CR-21, CR-22, and CR-24) formed at the earliest times of their host galaxy, essentially as part of the old bulge formation. Others (CR-11, CR-12, CR-18, and CR-20) formed later, with their counterrotating stars born during one or more bursts triggered by satellite interactions (but without significant stellar accretion). A particular case (CR-14) shows a minor contribution from accreted stars in addition to in situ star formation. The CRD of CR-16 is the remnant of a pre-disk stellar population formed prior to the development of a dynamically cold disk, while the CRD of CR-13 originated from the accretion of misaligned gas that fueled central star formation.

From these results, we can broadly categorize the origins of compact in situ CRDs into three scenarios: (i) formation at early times together with the galaxy’s bulge, (ii) interaction-induced star formation without substantial stellar accretion, (iii) peculiar cases associated with smooth misaligned cold gas accretion and pre-stellar disk irregular star formation. The first two mechanisms are the most frequent. In practice, some CRDs may involve a combination of these mechanisms. Overall, however, the common theme is that the compact in situ CRDs were built by star formation within the galaxy and often triggered or influenced by interactions rather than by direct deposition of stars from mergers.

\subsection{Compact ex situ CRDs}
\label{sec:compactex situCRD}

This subtype of CRD is compact in extent but contains a significant fraction of ex situ stellar particles. By definition, compact ex situ CRDs have more than 20\% of their counterrotating stars accreted from satellite galaxies. We identified seven galaxies in this category (CR-14, discussed earlier, falls just below the 20\% threshold). These galaxies are marked with red rhombus symbols in Figures \ref{fig:classificationtypes} and \ref{fig:ages_CRDs} (see also Appendix Figures \ref{fig:CRDs_all_appendix} and \ref{fig:CRDs_CRD_appendix}).

As shown on Fig. \ref{fig:assemblyinfo_cr}, past mergers are of great importance in the assembly of compact ex situ CRDs. In several cases, a single satellite merger provided the bulk of the accreted stellar particles for the CRD. For example, CR-4, CR-6, CR-7, CR-15, CR-19, and CR-23 each have one dominant progenitor galaxy contributing the majority of the ex situ component in the CRD. Only one case, CR-2, has two significant contributors to its ex situ stellar mass. Often, these mergers also enhanced in situ star formation in the host. As a result, the infalling satellites not only donated stars but also triggered new star formation in retrograde orbits, as evidenced by the SFHs. Indeed, Fig. \ref{fig:ages_CRDs} shows clear star formation peaks coincident with the merger times for these galaxies. We recall that, in all panels, the most significant contributor is shown with a highlighted arrow.

    \begin{figure}[t]
        \centering
        \includegraphics[width=3.2in]{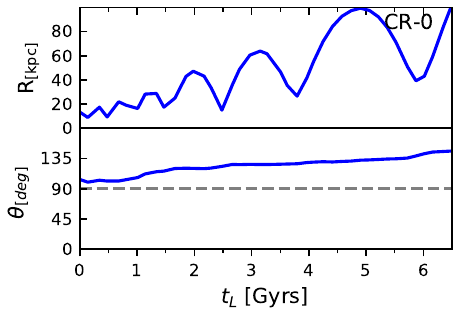}
        \caption{Orbit of the satellite that continues its interaction with CR-0 at z=0. We followed the evolution of the satellite orbit to a lookback time of 6 Gyr. Top panel: Evolution of the satellite galactocentric distance from the host. Bottom panel: Evolution of the angle between the angular momentum of the disk and the satellite's orbital angular momentum. Orbits near $180^\circ$ are counterrotating, while those near $0^\circ$ are corotating.}
        \label{fig:sat_info_CRDs}
    \end{figure}

Notably, some of the most massive satellite interactions in our entire sample are associated with compact ex situ CRDs. In particular, CR-2, CR-4, CR-6, and CR-7 experienced some of the largest mass merger events (in terms of total satellite-to-host mass ratio) of all the galaxies with CRDs (mass ratio > 40\%). These massive satellites are strongly affected by dynamical friction, and they rapidly sink to the host galactic center. The fact that these major interactions resulted in relatively compact coplanar CRDs (confined to the bulge region) indicates that the stars brought in by the merger and those induced by associated star-formation bursts remained concentrated in the central regions of the galaxy. It is also worth noting that, as in the compact in situ cases, these CRDs contain an old and significant in situ component tracing back to the formation of the galaxy’s primordial inner regions.

In observational terms, compact CRDs such as those identified in TNG50 may be difficult to detect, as their small radial extent and the dominance of the central bulge light limits their visibility in standard kinematic analyses. Observationally confirmed CRDs, such as those reported by \citet{2022ApJ...926L..13B}, typically correspond to more extended structures, likely due to selection biases or resolution constraints.  According to our analysis, if a compact ex situ CRD were to be observed, its formation  would be strongly influenced by one or two significant merger events. In all eight cases of this subtype in our sample, an interaction or merger can be linked to the emergence of the counterrotating component. While in simulations the ex situ mass fraction provides a clear discriminator of the CRD formation channel, such a quantity is inaccessible observationally. In practice, distinguishing in situ CRDs from ex situ CRDs would involve reliance on indirect indicators such as differences in stellar age, metallicity, or $\alpha$-element abundance relative to the host disk. In addition, compact CRDs could still produce detectable features in the stellar velocity dispersion field, such as double peaks (i.e., 2$\sigma$ or $\sigma$-elongated profiles), which do not require spatial separation of the corotating and counterrotating components. We defer the analysis of potential observational evidence of compact CRDs to a follow-up work.

\subsection{Extended in situ CRDs}
\label{sec:extendedin situCRD}

Extended CRDs are the minority in our TNG50 sample. By definition, these systems have  $R_{95}^{\rm CRD}/R_{\rm opt} > 0.5$, indicating that their counterrotating stellar disks extend beyond half the optical radius of the host galaxy. Among the 26 identified CRDs, only five meet this criterion. All five of these are predominantly of in situ origin and are marked with orange circle symbols in Fig. \ref{fig:classificationtypes} and \ref{fig:ages_CRDs} (see also Fig. \ref{fig:CRDs_all_appendix} and \ref{fig:CRDs_CRD_appendix}).

The extended in situ CRDs, CR-0, CR-3, CR-5, CR-10, and CR-17, formed the bulk of their counterrotating stars within the main galaxy, with relatively low ex situ stellar mass fractions. Specifically, approximately $11 -17\%$ of the CRD stellar mass of CR-0, CR-3, CR-5, and CR-10 was contributed by satellites, while CR-17 is almost entirely in situ, with less than $2\%$ of ex situ material. Despite their in situ dominance, all five systems show clear signs that interactions played a pivotal role in their formation, either by supplying retrograde gas or perturbing the host. As illustrated in Fig. \ref{fig:ages_CRDs}, which traces the stellar age distributions and merger events for each CRD, these galaxies experienced one or more significant interactions around the epoch of their main counterrotating star formation. Notably, and in contrast to the majority of CRDs in our sample, the stellar distribution in these five systems is significantly extended.

We show examples of this CRD type in Fig. \ref{fig:8subhaloes_cr}. Two peculiar cases are CR-0 and CR-5, which present a torus-like spatial distribution of the CRD component. Both present a concentration $R_{95}^{\rm CRD}/R_{50}^{\rm CRD} \leq 2$ (see Sec. \ref{section:CRDstructuralandassembly}). In the CR-0 case, the SFH of its CRD (Fig. \ref{fig:ages_CRDs}) shows at least three distinct peaks. The oldest peak, around 12 Gyr ago (lookback time), corresponds to the very early assembly of the galaxy and is associated with both gas accretion during these early times and a burst associated with a very significant merger, which we highlight with a green arrow in the figure. The next peak, near 9 Gyr ago, is associated with the infall of a second significant satellite (blue arrow in Fig. \ref{fig:ages_CRDs}). This interaction induced a burst of star formation that contributed to the CRD in the inner galactic region, as seen in the previous CRD type. The most recent rise in CR-0’s counterrotating star formation began $\sim 3$ Gyr ago and continues to the present day. Interestingly, CR-0 is experiencing an ongoing interaction with a massive satellite that had its first pericentric passage $\approx 4$ to 5 Gyr ago. The satellite has not fully merged at $z=0$ and continues to orbit the host. In Fig. \ref{fig:sat_info_CRDs}, we analyze the orbit of this infalling satellite. The top panel shows the time evolution of the satellite's galactocentric distance, while the bottom panel shows the evolution of the angle between the angular momentum of the disk and the satellites's orbital angular momentum. Orbits near $180^\circ$ are counterrotating, those near $90^\circ$ are polar, and those near $0^\circ$ are corotating. The figure clearly shows that the satellite is accreted in a counterotating orbit with respect to the main disk. The satellite is relatively gas rich, with a gas-to-total mass fraction at infall of 17\%. The gas stripped on this counterotating orbit fuels the more recent star formation episode of the CRD in the host galaxy.

\begin{figure}[t]
        \centering
        \includegraphics[width=3.2in]{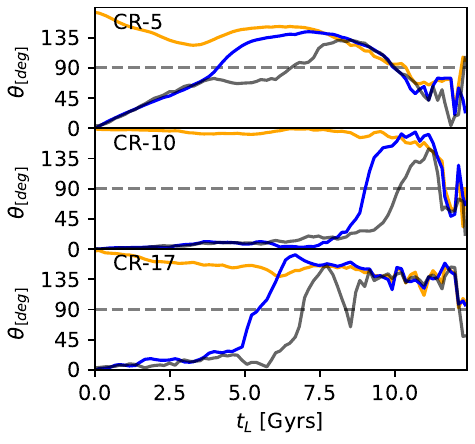}
        \caption{Time evolution of the angle between the angular momentum of the disk at z=0 and the disk (blue), CRD (orange), and gas within $R_{\rm opt}$ (black). Angles near $180^\circ$ are counterrotating, while those near $0^\circ$ are corotating in comparison to the present-day disk. Separations in angle between components represent a misalignment between them.}
        \label{fig:angle_components}
    \end{figure}

CR-5 also presents a torus-like CRD structure. Its star formation peak occurred around 9.5 Gyr ago and coincides with a major interaction involving a massive gas-rich satellite. This satellite had its first pericenter passage $\sim 11$ Gyr ago, and it deposited a significant amount of corotating gas at the outermost preexisting disk region, which subsequently formed the stellar torus-like structure. Interestingly, after the formation of this structure, CR-5 experienced a significant accretion of gas, counterrotating with respect to the preexisting structure, giving rise to the present-day disk. The rise of the younger dominant stellar disk is shown in the top panel of Fig. \ref{fig:angle_components}. In the figure, with blue lines we show the time evolution of the disk’s angular momentum vector orientation with respect to its orientation at the present-day, $J_{\rm disk}^{t_0}$. In addition, with an orange line we show  the time evolution of the CRD angular momentum vector orientation with respect to $J_{\rm disk}^{t_0}$. One can clearly see that the CRD and the overall disk remain well aligned until $ t_{L}= 7.5$ Gyr. After this time, both components become progressively more misaligned. This is due to the later accretion of misaligned gas, as shown by the black line, which follows the time evolution of the gas angular momentum vector orientation, $J_{\rm gas}$, with respect to $J_{\rm disk}^{t_0}$. To compute $J_{\rm gas}$, we considered at every snapshot all gas cells enclosed within a sphere of radius equal to the present-day $R_{\rm opt}$. This clearly indicates the significant accretion of misaligned cold gas within $4.5 \lesssim t_L \lesssim  7.5$ Gyr. Within this period of time, the increasing misalignment of the gas component is subsequently followed by what results in the present-day corotating disk.

Notably, CR-10 and CR-17 present a similar formation mechanism of the CRD component. In the case of CR-10, its CRD formed through multiple nearly simultaneous satellite interactions at an early time. Four different satellites (two with mass ratios over 25\%, and the other two with ratios of~12\%) interacted in a narrow time window with the host. We indicate the most massive interacting satellites with red arrows in Fig. \ref{fig:ages_CRDs}. A central burst of star formation associated with these interactions contributed to both the corotating and counterotating components of the final disk. These  satellites also contributed stellar particles and an extended gas component that gave rise to the present-day CRD. Interestingly, as shown in the middle panel of Fig. \ref{fig:angle_components}, the present-day CRD arises as a corotating component. However, the posterior accretion of misaligned gas starting at $t_{L} \approx 11.5$ Gyr gives rise to a new rotating component that, at the present day, represents the corotating disk. In contrast, CR-17 CRD formed almost entirely in situ (only ~2\% ex situ) within two early bursts of star formation: a first episode at a lookback time of $t_{L} \approx 10$ Gyr and a second episode 7 Gyr ago. We note that during both episodes of formation in the CRD, the star-formation activity of the present-day corotating component was almost negligible. Indeed, as shown in Fig.~\ref{fig:ages_CRDs}, the present-day corotating disk starts to grow in mass after $t_L \approx 7$ Gyr. In the bottom panel of Fig. \ref{fig:angle_components}, it can be clearly seen that the CRD is born as a corotating component. However, after $t_L \approx 7.5$ Gyr, significant accretion of counterotating gas gives rise to the present-day corotating disk.

Finally, CR-3 is an outlier in several aspects. It is the closest extended CRD to being ex situ; about 17\% of its stellar mass is directly accreted from satellites. CR-3 also extends unusually far into the outer disk (high $R_{95}^{\rm CRD}/R_{\rm opt}$). Its formation history (Fig. \ref{fig:ages_CRDs}) shows a prominent star formation peak around 8 Gyr ago, coinciding with the infall of multiple significant satellites in rapid succession. Two of these satellites were fairly massive ($\approx$ 40\% mass ratio). The ex situ stellar component of this CRD originates from these satellites. The interactions not only deposited stars into the CRD but also funneled gas into the host, which subsequently formed stars on retrograde orbits (hence the substantial in situ contribution).

In summary, all four extended in situ CRDs can trace their formation to interactions: CR-5 and CR-10 each formed during a major burst of multiple simultaneous minor mergers; the CRD of CR-0 grew through a combination of early accretion and an ongoing major interaction; and the CRD of CR-17 was built via a single low mass merger. However, the reasons for counterrotation vary. For CR-0, it is due to the ongoing accretion of the infalling satellite in a counterrotating orbit, whereas CR-5, CR-10, and CR-17 all have later accretion of misaligned gas that gives rise to a younger stellar disk that ends up dominating over the older CRD. Finally, CR-3 is mainly associated with the simultaneous accretion of significant satellites that contributed both stars and gas in counterotating orbits. The accretion of counterotating gas that gives rise to some of these CRDs will be studied in detail in a follow-up article.

\section{Discussion}
\label{sec:discussion}
Our study, based on fully cosmological simulations, indicates that CRDs are rare but present in $\lesssim 10\%$ of MW-mass galaxies at $z=0$, and that only $\lesssim 1\%$ host an extended CRD. These CRDs comprise only a small fraction of their hosts’ stellar mass (typically $\lesssim 2\%$) and tend to be spatially compact (confined well within the optical radius). This is opposed to lenticular and early-type disk galaxies where the counter-rotating components found in them tend to be of a higher mass fraction and are also more common in simulations (\citealt{2021MNRAS.500.3870K}, \citealt{2021MNRAS.503..726L}) and in observations \citep{2025ApJS..281...19G}, where coplanar as well as non-coplanar counterrotating disks are considered. A key result of our work is that all but one of the CRDs we find are dominated by in situ star formation, with the majority of stars in each CRD having formed within the galaxy’s potential well rather than being accreted. This is true even in cases with a non-negligible ex situ stellar contribution. In line with this, most CRDs have distinctly timed star formation episodes in their history – sharp bursts that coincide with some perturbation. This is in agreement with observations, where counterrotating stellar disks are a rare occurrence, encompassing only $\approx 1\%$ of galaxies with ionized gas in MaNGA (\citealt{2022ApJ...926L..13B}, \citealt{2022MNRAS.511..139B}). Interestingly, we find that the simulated CRDs are on average older than their present-day corotating stellar disks. This is in  contrast to observational studies, where counterrotating stellar disks are mostly found to be younger than the main stellar disk (\citealt{2014A&A...570A..79P}, \citealt{2025ApJS..281...19G}). We also find that galaxy models with CRDs typically have the gas and the younger stellar disk component (typically dominant) corotating with respect to each other. This in agreement with observations of counterrotating stellar disks in late-type galaxies (\citealt{2022ApJ...926L..13B}, \citealt{2024ApJ...973...29B}, \citealt{2025ApJS..281...19G}).

Our analysis revealed multiple formation channels for CRDs with varying degrees of external influence. The simulated galaxies exhibit a spectrum of CRD types, from significant merger-driven structures to more internal early star forming episodes. For example, one notable CRD involves an extended retrograde disk (CR-3) built primarily by the accretion of multiple satellite galaxies in quick succession. This event contributed both stellar particles and a significant amount of gas mass in retrograde orbits, producing one of the most extended CRDs in the sample. 
There are four other extended in situ CRDs that despite having nearly all their counterrotating stars formed in situ, can each be traced to at least one significant encounter that induced their formation. One of these cases is currently forming stars in its CRD as a result of the prolonged accretion of a satellite on a retrograde orbit. In contrast, the other three older CRDs are remnants of a previous accretion of misaligned gas that gave rise to the currently dominant younger corotating stellar disk. Episodes of gas misalignment giving rise to a younger stellar disk have been observed in past studies (\citealt{2004A&A...424..447P}; \citealt{2007A&A...463..883V}; \citealt{2025ApJS..281...19G}).
Similarly, we identified compact ex situ CRDs (eight cases), i.e., compact and retrograde disk components with a significant ex situ contribution. In general, these ex situ star particles are associated with one significant progenitor. The associated interactions often enhanced in situ star formation in the inner regions of the host (\citealt{1996ApJ...471..115B}, \citealt{2002MNRAS.333..327T}, \citealt{2010ApJ...710L.156R}, \citealt{2018MNRAS.479.3381B}), contributing to both the overall and the CR stellar distribution. 
Finally, the most common variety of CRDs in TNG50 are compact in situ CRDs. We identified 14 cases, and they are diverse in origin: Roughly half formed at very early times, coeval with the galaxy’s initial bulge formation and without any clear external trigger. A few others in this category show evidence of accretion events that induced bursts of in situ star formation within the central region that contributed to both the corotating and counterrotating components of the final stellar disk.

Despite the variety of formation channels, the dominance of the in situ stars indicates that gas with a misaligned (retrograde) angular momentum was accreted into the galaxy’s disk along its history. In some cases, this can be directly associated with interactions. Indeed, for the merger-induced CRDs, infalling satellites delivered not just stars but also fresh gas that was subsequently funneled into the host on retrograde orbits. In other cases, we found that misaligned gas accretion takes place after the present-day CRD formation, resulting in  a younger stellar disk that ends up dominating over the older CRD. The mechanism behind such accretion will be further studied in a follow-up work.

In summary, nearly all (all but one) simulated CRDs are dominated by stars formed in situ, even in systems shaped by mergers. Importantly, ‘‘in situ" here includes stars formed within the host from gas stripped or supplied by companions. Interactions often play a role in triggering or shaping the CRD. However, they are not strictly required in all cases. A small fraction of CRDs, especially compact ones, appear to arise naturally during the early phases of galaxy formation and show no evidence of any recent accretion event. Thus, our results indicate that although most counterrotating stellar disks are dominated by in situ star formation, their emergence is frequently linked to external processes. Interactions and the accretion of misaligned gas are common triggers that supply the conditions for retrograde star formation.

\section{Summary and conclusions}
\label{sec:conclusions}
In this work we have identified and analyzed the formation mechanisms of counterrotating stellar disks in galaxies selected from the Illustris TNG50 simulation. We limited our sample to only MW-mass galaxies, ranging in total mass from $10^{11.5}$ to $10^{12.5} $M$_\odot$. We characterized these counterrotating components by selecting particles with circularities of less than $-0.7$ and located within the spatial extension of the stellar disk. Within the sample of 260 MW-mass galaxies, we identified 26 cases with a counterrotating stellar disk. Their CRDs contribute to at least 1\% of the stellar mass of the galaxy. Our main results are summarized as follows:

\begin{itemize}
    
    \item Statistics: In the considered sample, CRDs are rare. Only $\approx 10\%$  of our sample hosts a significant CRD when adopting a threshold of $M_{\rm CRD}/M_{\rm Disk} \ge 0.01$. As expected in late-type galaxies, CRDs represent a minor fraction of the overall stellar disk mass.

    \item Structure: CRDs from our sample are predominantly compact. Quantitatively, \(\approx 88\%\) have \(R^{\rm CRD}_{50}<0.25\,R_{\rm opt}\), and the distribution of the sizes peaks at a median of \(R^{\rm CRD}_{95}/R_{\rm opt}\sim 0.37\), although a minority are more extended. Notably, three systems display torus-like morphologies.

    \item Stellar origin: Nearly all the CRDs in our sample are in situ dominated. The definition we used includes stars formed inside the host from gas that may have been stripped from satellites. Ex situ contributions are typically small. Approximately 73\% of the CRDs have ex situ fractions \(<0.2\), and the median ex situ fraction is \(\approx 0.06\).

    \item Star formation histories and triggers: CRDs from the analyzed sample are generally older than their corotating counterparts and exhibit bursty SFHs whose peaks frequently coincide with satellite pericenters or interactions. Thus, external perturbations often trigger star formation. These episodes may occur in retrograde or prograde orbits, which are subsequently transformed into a CRD through the later accretion of cold gas, which gives rise to the present-day disk.

    \item Formation channels: A variety of formation mechanisms appear in our sample of CRDs. Compact in situ CRDs arise through multiple pathways: (i) early formation alongside bulge assembly; (ii) interaction-induced in situ bursts; and (iii) rarer cases linked to smooth misaligned gas accretion. Extended in situ CRDs, although fewer, can usually be traced to interactions and late accretion of misaligned gas that gives rise to a younger dominant corotating stellar disk. ex situ rich cases typically originate from one dominant (occasionally two) progenitor(s); these mergers often coincide with enhanced in situ star formation in retrograde orbits.

\end{itemize}

In a follow-up work we will characterize the mechanism that drives the late accretion of misaligned cold gas and preferentially gives rise to extended CRDs. We will also explore whether the identification of compact in situ CRDs can be achieved through the characterization of stellar population abundances.

\begin{acknowledgements}
We thank the anonymous referee for the insightful comments and constructive suggestions that helped improve this work.
MCB, FAG, AD and PBT acknowledges support from the ANID BASAL project FB210003. FAG acknowledges support from  the ANID FONDECYT Regular grant 1251493. FAG, AD and PBT acknowledges support from the HORIZON MSCA-2021-SE-01 Research and Innovation Programme under the Marie Sklodowska-Curie grant agreement number 101086388. AD acknowledges support from the FONDECYT Postdoctorado No. 3250558. PBT acknowledges partial support from Fondecyt Regular ANID 1240465. The authors acknowledge the use of ChatGPT to improve the clarity and readability of the manuscript.
\end{acknowledgements}

\bibliographystyle{aa}
\bibliography{references.bib}

@ARTICLE{2025ApJS..281...19G,
       author = {{Gasymov}, Damir and {Katkov}, Ivan Yu. and {Rubtsov}, Evgenii V. and {Saburova}, Anna S. and {Kniazev}, Alexei Yu. and {Gelfand}, Joseph D. and {Sil'chenko}, Olga K. and {Chilingarian}, Igor V. and {Moiseev}, Alexei V. and {Kasparova}, Anastasia V. and {Zasov}, Anatoly V.},
        title = "{Stellar CoRGI in MaNGA: Stellar Counterrotating Galaxies Identified in the MaNGA Survey}",
      journal = {\apjs},
         year = 2025,
        month = nov,
       volume = {281},
       number = {1},
          eid = {19},
        pages = {19},
          doi = {10.3847/1538-4365/ae0590},
archivePrefix = {arXiv},
       eprint = {2504.02925},
 primaryClass = {astro-ph.GA},
       adsurl = {https://ui.adsabs.harvard.edu/abs/2025ApJS..281...19G}
}

@ARTICLE{2025A&A...696A..45P,
       author = {{Peirani}, S{\'e}bastien and {Suto}, Yasushi and {Han}, Seongbong and {Yi}, Sukyoung K. and {Dubois}, Yohan and {Kraljic}, Katarina and {Park}, Minjung and {Pichon}, Christophe},
        title = "{Dissecting the formation of gas-versus-star counter-rotating galaxies from the NewHorizon simulation}",
      journal = {\aap},
         year = 2025,
        month = apr,
       volume = {696},
          eid = {A45},
        pages = {A45},
          doi = {10.1051/0004-6361/202453577},
archivePrefix = {arXiv},
       eprint = {2502.17902},
 primaryClass = {astro-ph.GA},
       adsurl = {https://ui.adsabs.harvard.edu/abs/2025A&A...696A..45P}
}

@ARTICLE{2024ApJ...973...29B,
       author = {{Bao}, Min and {Zhao}, Wenlong and {Yuan}, Qirong},
        title = "{Different Influence of Gas Accretion on the Evolution of Star-forming and Non-star-forming Galaxies}",
      journal = {\apj},
         year = 2024,
        month = sep,
       volume = {973},
       number = {1},
          eid = {29},
        pages = {29},
          doi = {10.3847/1538-4357/ad6441},
archivePrefix = {arXiv},
       eprint = {2407.11340},
 primaryClass = {astro-ph.GA},
       adsurl = {https://ui.adsabs.harvard.edu/abs/2024ApJ...973...29B}
}

@ARTICLE{2024MNRAS.528.2326S,
       author = {{Santucci}, Giulia and {Lagos}, Claudia Del P. and {Harborne}, Katherine E. and {Ludlow}, Aaron and {Proctor}, Katy L. and {Foster}, Caroline and {McDermid}, Richard and {Poci}, Adriano and {Thater}, Sabine and {van de Ven}, Glenn and {Zhu}, Ling and {Walo Mart{\'\i}n}, Daniel},
        title = "{The distribution of stellar orbits in EAGLE galaxies - the effect of mergers, gas accretion, and secular evolution}",
      journal = {\mnras},
         year = 2024,
        month = feb,
       volume = {528},
       number = {2},
        pages = {2326-2345},
          doi = {10.1093/mnras/stae113},
archivePrefix = {arXiv},
       eprint = {2310.06885},
 primaryClass = {astro-ph.GA},
       adsurl = {https://ui.adsabs.harvard.edu/abs/2024MNRAS.528.2326S}
}

@ARTICLE{2023MNRAS.526..567D,
       author = {{Dolfi}, Arianna and {G{\'o}mez}, Facundo A. and {Monachesi}, Antonela and {Varela-Lavin}, Silvio and {Tissera}, Patricia B. and {Sif{\'o}n}, Crist{\'o}bal and {Galaz}, Gaspar},
        title = "{Lopsidedness as a tracer of early galactic assembly history}",
      journal = {\mnras},
         year = 2023,
        month = nov,
       volume = {526},
       number = {1},
        pages = {567-584},
          doi = {10.1093/mnras/stad2650},
archivePrefix = {arXiv},
       eprint = {2306.04639},
 primaryClass = {astro-ph.GA},
       adsurl = {https://ui.adsabs.harvard.edu/abs/2023MNRAS.526..567D}
}

@ARTICLE{2023MNRAS.523.5853V,
       author = {{Varela-Lavin}, Silvio and {G{\'o}mez}, Facundo A. and {Tissera}, Patricia B. and {Besla}, Gurtina and {Garavito-Camargo}, Nicol{\'a}s and {Marinacci}, Federico and {Laporte}, Chervin F.~P.},
        title = "{Lopsided galaxies in a cosmological context: a new galaxy-halo connection}",
      journal = {\mnras},
         year = 2023,
        month = aug,
       volume = {523},
       number = {4},
        pages = {5853-5868},
          doi = {10.1093/mnras/stad1724},
archivePrefix = {arXiv},
       eprint = {2211.16577},
 primaryClass = {astro-ph.GA},
       adsurl = {https://ui.adsabs.harvard.edu/abs/2023MNRAS.523.5853V}
}

@ARTICLE{2023AstL...49..229S,
       author = {{Sil'chenko}, O.~K. and {Moiseev}, A.~V. and {Oparin}, D.~V. and {Zlydneva}, D.~V. and {Kozlova}, D.~V.},
        title = "{Counter-Rotating Gaseous Disk and Star Formation in the S0 Galaxy NGC 934}",
      journal = {Astronomy Letters},
         year = 2023,
        month = may,
       volume = {49},
       number = {5},
        pages = {229-239},
          doi = {10.1134/S1063773723050043},
       adsurl = {https://ui.adsabs.harvard.edu/abs/2023AstL...49..229S}
}

@ARTICLE{2022MNRAS.511..139B,
       author = {{Bevacqua}, Davide and {Cappellari}, Michele and {Pellegrini}, Silvia},
        title = "{SDSS-IV MaNGA: integral-field kinematics and stellar population of a sample of galaxies with counter-rotating stellar discs selected from about 4000 galaxies}",
      journal = {\mnras},
         year = 2022,
        month = mar,
       volume = {511},
       number = {1},
        pages = {139-157},
          doi = {10.1093/mnras/stab3732},
archivePrefix = {arXiv},
       eprint = {2107.09528},
 primaryClass = {astro-ph.GA},
       adsurl = {https://ui.adsabs.harvard.edu/abs/2022MNRAS.511..139B}
}

@ARTICLE{2022ApJ...926L..13B,
       author = {{Bao}, Min and {Chen}, Yanmei and {Zhu}, Pengpei and {Shi}, Yong and {Bizyaev}, Dmitry and {Zhu}, Ling and {Yang}, Meng and {Beom}, Minje and {Brownstein}, Joel R. and {Lane}, Richard R.},
        title = "{Different Formation Scenarios for Counterrotating Stellar Disks in Nearby Galaxies}",
      journal = {\apjl},
         year = 2022,
        month = feb,
       volume = {926},
       number = {2},
          eid = {L13},
        pages = {L13},
          doi = {10.3847/2041-8213/ac52ad},
archivePrefix = {arXiv},
       eprint = {2202.03848},
 primaryClass = {astro-ph.GA},
       adsurl = {https://ui.adsabs.harvard.edu/abs/2022ApJ...926L..13B}
}

@ARTICLE{2022MNRAS.509.1764J,
       author = {{Jagvaral}, Yesukhei and {Campbell}, Duncan and {Mandelbaum}, Rachel and {Rau}, Markus Michael},
        title = "{Probabilistic model for dynamic galaxy decomposition}",
      journal = {\mnras},
         year = 2022,
        month = jan,
       volume = {509},
       number = {2},
        pages = {1764-1778},
          doi = {10.1093/mnras/stab3104},
archivePrefix = {arXiv},
       eprint = {2105.02237},
 primaryClass = {astro-ph.GA},
       adsurl = {https://ui.adsabs.harvard.edu/abs/2022MNRAS.509.1764J}
}

@ARTICLE{2021A&A...654A..30R,
       author = {{Rubino}, M. and {Pizzella}, A. and {Morelli}, L. and {Coccato}, L. and {Portaluri}, E. and {Debattista}, V.~P. and {Corsini}, E.~M. and {Dalla Bont{\`a}}, E.},
        title = "{Detectability of large-scale counter-rotating stellar disks in galaxies with integral-field spectroscopy}",
      journal = {\aap},
         year = 2021,
        month = oct,
       volume = {654},
          eid = {A30},
        pages = {A30},
          doi = {10.1051/0004-6361/202140702},
archivePrefix = {arXiv},
       eprint = {2107.02226},
 primaryClass = {astro-ph.GA},
       adsurl = {https://ui.adsabs.harvard.edu/abs/2021A&A...654A..30R}
}

@ARTICLE{2021ApJS..254...27K,
       author = {{Khim}, Donghyeon J. and {Yi}, Sukyoung K. and {Pichon}, Christophe and {Dubois}, Yohan and {Devriendt}, Julien and {Choi}, Hoseung and {Bryant}, Julia J. and {Croom}, Scott M.},
        title = "{Star-Gas Misalignment in Galaxies. II. Origins Found from the Horizon-AGN Simulation}",
      journal = {\apjs},
         year = 2021,
        month = jun,
       volume = {254},
       number = {2},
          eid = {27},
        pages = {27},
          doi = {10.3847/1538-4365/abf043},
archivePrefix = {arXiv},
       eprint = {2012.04659},
 primaryClass = {astro-ph.GA},
       adsurl = {https://ui.adsabs.harvard.edu/abs/2021ApJS..254...27K}
}

@ARTICLE{2021MNRAS.503..726L,
       author = {{Lu}, Shengdong and {Xu}, Dandan and {Wang}, Yunchong and {Chen}, Yanmei and {Zhu}, Ling and {Mao}, Shude and {Springel}, Volker and {Wang}, Jing and {Vogelsberger}, Mark and {Hernquist}, Lars},
        title = "{Hot and counter-rotating star-forming disc galaxies in IllustrisTNG and their real-world counterparts}",
      journal = {\mnras},
         year = 2021,
        month = may,
       volume = {503},
       number = {1},
        pages = {726-742},
          doi = {10.1093/mnras/stab497},
archivePrefix = {arXiv},
       eprint = {2011.01949},
 primaryClass = {astro-ph.GA},
       adsurl = {https://ui.adsabs.harvard.edu/abs/2021MNRAS.503..726L}
}

@ARTICLE{2021MNRAS.500.3870K,
       author = {{Khoperskov}, Sergey and {Zinchenko}, Igor and {Avramov}, Branislav and {Khrapov}, Sergey and {Berczik}, Peter and {Saburova}, Anna and {Ishchenko}, Marina and {Khoperskov}, Alexander and {Pulsoni}, Claudia and {Venichenko}, Yulia and {Bizyaev}, Dmitry and {Moiseev}, Alexei},
        title = "{Extreme kinematic misalignment in IllustrisTNG galaxies: the origin, structure, and internal dynamics of galaxies with a large-scale counterrotation}",
      journal = {\mnras},
         year = 2021,
        month = jan,
       volume = {500},
       number = {3},
        pages = {3870-3888},
          doi = {10.1093/mnras/staa3330},
archivePrefix = {arXiv},
       eprint = {2010.11581},
 primaryClass = {astro-ph.GA},
       adsurl = {https://ui.adsabs.harvard.edu/abs/2021MNRAS.500.3870K}
}

@ARTICLE{2019MNRAS.490.3234N,
       author = {{Nelson}, Dylan and {Pillepich}, Annalisa and {Springel}, Volker and {Pakmor}, R{\"u}diger and {Weinberger}, Rainer and {Genel}, Shy and {Torrey}, Paul and {Vogelsberger}, Mark and {Marinacci}, Federico and {Hernquist}, Lars},
        title = "{First results from the TNG50 simulation: galactic outflows driven by supernovae and black hole feedback}",
      journal = {\mnras},
         year = 2019,
        month = dec,
       volume = {490},
       number = {3},
        pages = {3234-3261},
          doi = {10.1093/mnras/stz2306},
archivePrefix = {arXiv},
       eprint = {1902.05554},
 primaryClass = {astro-ph.GA},
       adsurl = {https://ui.adsabs.harvard.edu/abs/2019MNRAS.490.3234N}
}

@ARTICLE{2019MNRAS.490.3196P,
       author = {{Pillepich}, Annalisa and {Nelson}, Dylan and {Springel}, Volker and {Pakmor}, R{\"u}diger and {Torrey}, Paul and {Weinberger}, Rainer and {Vogelsberger}, Mark and {Marinacci}, Federico and {Genel}, Shy and {van der Wel}, Arjen and {Hernquist}, Lars},
        title = "{First results from the TNG50 simulation: the evolution of stellar and gaseous discs across cosmic time}",
      journal = {\mnras},
         year = 2019,
        month = dec,
       volume = {490},
       number = {3},
        pages = {3196-3233},
          doi = {10.1093/mnras/stz2338},
archivePrefix = {arXiv},
       eprint = {1902.05553},
 primaryClass = {astro-ph.GA},
       adsurl = {https://ui.adsabs.harvard.edu/abs/2019MNRAS.490.3196P}
}

@ARTICLE{2019ApJ...878..143S,
       author = {{Starkenburg}, Tjitske K. and {Sales}, Laura. V. and {Genel}, Shy and {Manzano-King}, Christina and {Canalizo}, Gabriela and {Hernquist}, Lars},
        title = "{On the Origin of Star-Gas Counterrotation in Low-mass Galaxies}",
      journal = {\apj},
         year = 2019,
        month = jun,
       volume = {878},
       number = {2},
          eid = {143},
        pages = {143},
          doi = {10.3847/1538-4357/ab2128},
archivePrefix = {arXiv},
       eprint = {1903.03627},
 primaryClass = {astro-ph.GA},
       adsurl = {https://ui.adsabs.harvard.edu/abs/2019ApJ...878..143S}
}

@ARTICLE{2019MNRAS.485.2589M,
       author = {{Monachesi}, Antonela and {G{\'o}mez}, Facundo A. and {Grand}, Robert J.~J. and {Simpson}, Christine M. and {Kauffmann}, Guinevere and {Bustamante}, Sebasti{\'a}n and {Marinacci}, Federico and {Pakmor}, R{\"u}diger and {Springel}, Volker and {Frenk}, Carlos S. and {White}, Simon D.~M. and {Tissera}, Patricia B.},
        title = "{The Auriga stellar haloes: connecting stellar population properties with accretion and merging history}",
      journal = {\mnras},
         year = 2019,
        month = may,
       volume = {485},
       number = {2},
        pages = {2589-2616},
          doi = {10.1093/mnras/stz538},
archivePrefix = {arXiv},
       eprint = {1804.07798},
 primaryClass = {astro-ph.GA},
       adsurl = {https://ui.adsabs.harvard.edu/abs/2019MNRAS.485.2589M}
}

@ARTICLE{2018MNRAS.480.5113M,
       author = {{Marinacci}, Federico and {Vogelsberger}, Mark and {Pakmor}, R{\"u}diger and {Torrey}, Paul and {Springel}, Volker and {Hernquist}, Lars and {Nelson}, Dylan and {Weinberger}, Rainer and {Pillepich}, Annalisa and {Naiman}, Jill and {Genel}, Shy},
        title = "{First results from the IllustrisTNG simulations: radio haloes and magnetic fields}",
      journal = {\mnras},
         year = 2018,
        month = nov,
       volume = {480},
       number = {4},
        pages = {5113-5139},
          doi = {10.1093/mnras/sty2206},
archivePrefix = {arXiv},
       eprint = {1707.03396},
 primaryClass = {astro-ph.CO},
       adsurl = {https://ui.adsabs.harvard.edu/abs/2018MNRAS.480.5113M}
}

@ARTICLE{2018MNRAS.480.2544R,
       author = {{Rowlands}, K. and {Heckman}, T. and {Wild}, V. and {Zakamska}, N.~L. and {Rodriguez-Gomez}, V. and {Barrera-Ballesteros}, J. and {Lotz}, J. and {Thilker}, D. and {Andrews}, B.~H. and {Boquien}, M. and {Brinkmann}, J. and {Brownstein}, J.~R. and {Hwang}, H.-C. and {Smethurst}, R.},
        title = "{SDSS-IV MaNGA: spatially resolved star formation histories and the connection to galaxy physical properties}",
      journal = {\mnras},
         year = 2018,
        month = oct,
       volume = {480},
       number = {2},
        pages = {2544-2561},
          doi = {10.1093/mnras/sty1916},
archivePrefix = {arXiv},
       eprint = {1807.06066},
 primaryClass = {astro-ph.GA},
       adsurl = {https://ui.adsabs.harvard.edu/abs/2018MNRAS.480.2544R}
}

@ARTICLE{2018MNRAS.479.3381B,
       author = {{Bustamante}, Sebasti{\'a}n and {Sparre}, Martin and {Springel}, Volker and {Grand}, Robert J.~J.},
        title = "{Merger-induced metallicity dilution in cosmological galaxy formation simulations}",
      journal = {\mnras},
         year = 2018,
        month = sep,
       volume = {479},
       number = {3},
        pages = {3381-3392},
          doi = {10.1093/mnras/sty1692},
archivePrefix = {arXiv},
       eprint = {1712.03250},
 primaryClass = {astro-ph.GA},
       adsurl = {https://ui.adsabs.harvard.edu/abs/2018MNRAS.479.3381B}
}

@ARTICLE{2018A&A...616A..22P,
       author = {{Pizzella}, A. and {Morelli}, L. and {Coccato}, L. and {Corsini}, E.~M. and {Dalla Bont{\`a}}, E. and {Fabricius}, M. and {Saglia}, R.~P.},
        title = "{Evidence for the formation of the young counter-rotating stellar disk from gas acquired by IC 719}",
      journal = {\aap},
         year = 2018,
        month = aug,
       volume = {616},
          eid = {A22},
        pages = {A22},
          doi = {10.1051/0004-6361/201731712},
archivePrefix = {arXiv},
       eprint = {1804.03569},
 primaryClass = {astro-ph.GA},
       adsurl = {https://ui.adsabs.harvard.edu/abs/2018A&A...616A..22P}
}

@ARTICLE{2018MNRAS.477.1206N,
       author = {{Naiman}, Jill P. and {Pillepich}, Annalisa and {Springel}, Volker and {Ramirez-Ruiz}, Enrico and {Torrey}, Paul and {Vogelsberger}, Mark and {Pakmor}, R{\"u}diger and {Nelson}, Dylan and {Marinacci}, Federico and {Hernquist}, Lars and {Weinberger}, Rainer and {Genel}, Shy},
        title = "{First results from the IllustrisTNG simulations: a tale of two elements - chemical evolution of magnesium and europium}",
      journal = {\mnras},
         year = 2018,
        month = jun,
       volume = {477},
       number = {1},
        pages = {1206-1224},
          doi = {10.1093/mnras/sty618},
archivePrefix = {arXiv},
       eprint = {1707.03401},
 primaryClass = {astro-ph.GA},
       adsurl = {https://ui.adsabs.harvard.edu/abs/2018MNRAS.477.1206N}
}

@ARTICLE{2018MNRAS.475..624N,
       author = {{Nelson}, Dylan and {Pillepich}, Annalisa and {Springel}, Volker and {Weinberger}, Rainer and {Hernquist}, Lars and {Pakmor}, R{\"u}diger and {Genel}, Shy and {Torrey}, Paul and {Vogelsberger}, Mark and {Kauffmann}, Guinevere and {Marinacci}, Federico and {Naiman}, Jill},
        title = "{First results from the IllustrisTNG simulations: the galaxy colour bimodality}",
      journal = {\mnras},
         year = 2018,
        month = mar,
       volume = {475},
       number = {1},
        pages = {624-647},
          doi = {10.1093/mnras/stx3040},
archivePrefix = {arXiv},
       eprint = {1707.03395},
 primaryClass = {astro-ph.GA},
       adsurl = {https://ui.adsabs.harvard.edu/abs/2018MNRAS.475..624N}
}

@ARTICLE{2018MNRAS.475..648P,
       author = {{Pillepich}, Annalisa and {Nelson}, Dylan and {Hernquist}, Lars and {Springel}, Volker and {Pakmor}, R{\"u}diger and {Torrey}, Paul and {Weinberger}, Rainer and {Genel}, Shy and {Naiman}, Jill P. and {Marinacci}, Federico and {Vogelsberger}, Mark},
        title = "{First results from the IllustrisTNG simulations: the stellar mass content of groups and clusters of galaxies}",
      journal = {\mnras},
         year = 2018,
        month = mar,
       volume = {475},
       number = {1},
        pages = {648-675},
          doi = {10.1093/mnras/stx3112},
archivePrefix = {arXiv},
       eprint = {1707.03406},
 primaryClass = {astro-ph.GA},
       adsurl = {https://ui.adsabs.harvard.edu/abs/2018MNRAS.475..648P}
}

@ARTICLE{2018MNRAS.475..676S,
       author = {{Springel}, Volker and {Pakmor}, R{\"u}diger and {Pillepich}, Annalisa and {Weinberger}, Rainer and {Nelson}, Dylan and {Hernquist}, Lars and {Vogelsberger}, Mark and {Genel}, Shy and {Torrey}, Paul and {Marinacci}, Federico and {Naiman}, Jill},
        title = "{First results from the IllustrisTNG simulations: matter and galaxy clustering}",
      journal = {\mnras},
         year = 2018,
        month = mar,
       volume = {475},
       number = {1},
        pages = {676-698},
          doi = {10.1093/mnras/stx3304},
archivePrefix = {arXiv},
       eprint = {1707.03397},
 primaryClass = {astro-ph.GA},
       adsurl = {https://ui.adsabs.harvard.edu/abs/2018MNRAS.475..676S}
}

@ARTICLE{2017MNRAS.472.3722G,
       author = {{G{\'o}mez}, Facundo A. and {Grand}, Robert J.~J. and {Monachesi}, Antonela and {White}, Simon D.~M. and {Bustamante}, Sebastian and {Marinacci}, Federico and {Pakmor}, R{\"u}diger and {Simpson}, Christine M. and {Springel}, Volker and {Frenk}, Carlos S.},
        title = "{Lessons from the Auriga discs: the hunt for the Milky Way's ex situ disc is not yet over}",
      journal = {\mnras},
         year = 2017,
        month = dec,
       volume = {472},
       number = {3},
        pages = {3722-3733},
          doi = {10.1093/mnras/stx2149},
archivePrefix = {arXiv},
       eprint = {1704.08261},
 primaryClass = {astro-ph.GA},
       adsurl = {https://ui.adsabs.harvard.edu/abs/2017MNRAS.472.3722G}
}

@ARTICLE{2017MNRAS.472.4133G,
       author = {{Gargiulo}, I.~D. and {Cora}, S.~A. and {Vega-Mart{\'\i}nez}, C.~A. and {Gonzalez}, O.~A. and {Zoccali}, M. and {Gonz{\'a}lez}, R. and {Ruiz}, A.~N. and {Padilla}, N.~D.},
        title = "{Stellar Populations in a semi-analytic model I: Bulges of Milky Way-like galaxies}",
      journal = {\mnras},
         year = 2017,
        month = dec,
       volume = {472},
       number = {4},
        pages = {4133-4143},
          doi = {10.1093/mnras/stx2188},
archivePrefix = {arXiv},
       eprint = {1709.04521},
 primaryClass = {astro-ph.GA},
       adsurl = {https://ui.adsabs.harvard.edu/abs/2017MNRAS.472.4133G}
}

@ARTICLE{2017A&A...600A..76M,
       author = {{Morelli}, L. and {Pizzella}, A. and {Coccato}, L. and {Corsini}, E.~M. and {Dalla Bont{\`a}}, E. and {Buson}, L.~M. and {Ivanov}, V.~D. and {Pagotto}, I. and {Pompei}, E. and {Rocco}, M.},
        title = "{Kinematic and stellar population properties of the counter-rotating components in the S0 galaxy NGC 1366}",
      journal = {\aap},
         year = 2017,
        month = apr,
       volume = {600},
          eid = {A76},
        pages = {A76},
          doi = {10.1051/0004-6361/201630046},
archivePrefix = {arXiv},
       eprint = {1701.07631},
 primaryClass = {astro-ph.GA},
       adsurl = {https://ui.adsabs.harvard.edu/abs/2017A&A...600A..76M}
}

@ARTICLE{2017MNRAS.464.4789M,
       author = {{Mitzkus}, Martin and {Cappellari}, Michele and {Walcher}, C. Jakob},
        title = "{Dominant dark matter and a counter-rotating disc: MUSE view of the low-luminosity S0 galaxy NGC 5102}",
      journal = {\mnras},
         year = 2017,
        month = feb,
       volume = {464},
       number = {4},
        pages = {4789-4806},
          doi = {10.1093/mnras/stw2677},
archivePrefix = {arXiv},
       eprint = {1610.04516},
 primaryClass = {astro-ph.GA},
       adsurl = {https://ui.adsabs.harvard.edu/abs/2017MNRAS.464.4789M}
}

@ARTICLE{2016A&A...594A..13P,
       author = {{Planck Collaboration} and {Ade}, P.~A.~R. and {Aghanim}, N. and {Arnaud}, M. and {Ashdown}, M. and {Aumont}, J. and {Baccigalupi}, C. and {Banday}, A.~J. and {Barreiro}, R.~B. and {Bartlett}, J.~G. and {Bartolo}, N. and {Battaner}, E. and {Battye}, R. and {Benabed}, K. and {Beno{\^\i}t}, A. and {Benoit-L{\'e}vy}, A. and {Bernard}, J.-P. and {Bersanelli}, M. and {Bielewicz}, P. and {Bock}, J.~J. and {Bonaldi}, A. and {Bonavera}, L. and {Bond}, J.~R. and {Borrill}, J. and {Bouchet}, F.~R. and {Boulanger}, F. and {Bucher}, M. and {Burigana}, C. and {Butler}, R.~C. and {Calabrese}, E. and {Cardoso}, J.-F. and {Catalano}, A. and {Challinor}, A. and {Chamballu}, A. and {Chary}, R.-R. and {Chiang}, H.~C. and {Chluba}, J. and {Christensen}, P.~R. and {Church}, S. and {Clements}, D.~L. and {Colombi}, S. and {Colombo}, L.~P.~L. and {Combet}, C. and {Coulais}, A. and {Crill}, B.~P. and {Curto}, A. and {Cuttaia}, F. and {Danese}, L. and {Davies}, R.~D. and {Davis}, R.~J. and {de Bernardis}, P. and {de Rosa}, A. and {de Zotti}, G. and {Delabrouille}, J. and {D{\'e}sert}, F.-X. and {Di Valentino}, E. and {Dickinson}, C. and {Diego}, J.~M. and {Dolag}, K. and {Dole}, H. and {Donzelli}, S. and {Dor{\'e}}, O. and {Douspis}, M. and {Ducout}, A. and {Dunkley}, J. and {Dupac}, X. and {Efstathiou}, G. and {Elsner}, F. and {En{\ss}lin}, T.~A. and {Eriksen}, H.~K. and {Farhang}, M. and {Fergusson}, J. and {Finelli}, F. and {Forni}, O. and {Frailis}, M. and {Fraisse}, A.~A. and {Franceschi}, E. and {Frejsel}, A. and {Galeotta}, S. and {Galli}, S. and {Ganga}, K. and {Gauthier}, C. and {Gerbino}, M. and {Ghosh}, T. and {Giard}, M. and {Giraud-H{\'e}raud}, Y. and {Giusarma}, E. and {Gjerl{\o}w}, E. and {Gonz{\'a}lez-Nuevo}, J. and {G{\'o}rski}, K.~M. and {Gratton}, S. and {Gregorio}, A. and {Gruppuso}, A. and {Gudmundsson}, J.~E. and {Hamann}, J. and {Hansen}, F.~K. and {Hanson}, D. and {Harrison}, D.~L. and {Helou}, G. and {Henrot-Versill{\'e}}, S. and {Hern{\'a}ndez-Monteagudo}, C. and {Herranz}, D. and {Hildebrandt}, S.~R. and {Hivon}, E. and {Hobson}, M. and {Holmes}, W.~A. and {Hornstrup}, A. and {Hovest}, W. and {Huang}, Z. and {Huffenberger}, K.~M. and {Hurier}, G. and {Jaffe}, A.~H. and {Jaffe}, T.~R. and {Jones}, W.~C. and {Juvela}, M. and {Keih{\"a}nen}, E. and {Keskitalo}, R. and {Kisner}, T.~S. and {Kneissl}, R. and {Knoche}, J. and {Knox}, L. and {Kunz}, M. and {Kurki-Suonio}, H. and {Lagache}, G. and {L{\"a}hteenm{\"a}ki}, A. and {Lamarre}, J.-M. and {Lasenby}, A. and {Lattanzi}, M. and {Lawrence}, C.~R. and {Leahy}, J.~P. and {Leonardi}, R. and {Lesgourgues}, J. and {Levrier}, F. and {Lewis}, A. and {Liguori}, M. and {Lilje}, P.~B. and {Linden-V{\o}rnle}, M. and {L{\'o}pez-Caniego}, M. and {Lubin}, P.~M. and {Mac{\'\i}as-P{\'e}rez}, J.~F. and {Maggio}, G. and {Maino}, D. and {Mandolesi}, N. and {Mangilli}, A. and {Marchini}, A. and {Maris}, M. and {Martin}, P.~G. and {Martinelli}, M. and {Mart{\'\i}nez-Gonz{\'a}lez}, E. and {Masi}, S. and {Matarrese}, S. and {McGehee}, P. and {Meinhold}, P.~R. and {Melchiorri}, A. and {Melin}, J.-B. and {Mendes}, L. and {Mennella}, A. and {Migliaccio}, M. and {Millea}, M. and {Mitra}, S. and {Miville-Desch{\^e}nes}, M.-A. and {Moneti}, A. and {Montier}, L. and {Morgante}, G. and {Mortlock}, D. and {Moss}, A. and {Munshi}, D. and {Murphy}, J.~A. and {Naselsky}, P. and {Nati}, F. and {Natoli}, P. and {Netterfield}, C.~B. and {N{\o}rgaard-Nielsen}, H.~U. and {Noviello}, F. and {Novikov}, D. and {Novikov}, I. and {Oxborrow}, C.~A. and {Paci}, F. and {Pagano}, L. and {Pajot}, F. and {Paladini}, R. and {Paoletti}, D. and {Partridge}, B. and {Pasian}, F. and {Patanchon}, G. and {Pearson}, T.~J. and {Perdereau}, O. and {Perotto}, L. and {Perrotta}, F. and {Pettorino}, V. and {Piacentini}, F. and {Piat}, M. and {Pierpaoli}, E. and {Pietrobon}, D. and {Plaszczynski}, S. and {Pointecouteau}, E. and {Polenta}, G. and {Popa}, L. and {Pratt}, G.~W. and {Pr{\'e}zeau}, G.},
        title = "{Planck 2015 results. XIII. Cosmological parameters}",
      journal = {\aap},
         year = 2016,
        month = sep,
       volume = {594},
          eid = {A13},
        pages = {A13},
          doi = {10.1051/0004-6361/201525830},
archivePrefix = {arXiv},
       eprint = {1502.01589},
 primaryClass = {astro-ph.CO},
       adsurl = {https://ui.adsabs.harvard.edu/abs/2016A&A...594A..13P}
}

@ARTICLE{2016MNRAS.461.2068K,
       author = {{Katkov}, Ivan Yu. and {Sil'chenko}, Olga K. and {Chilingarian}, Igor V. and {Uklein}, Roman I. and {Egorov}, Oleg V.},
        title = "{Stellar counter-rotation in lenticular galaxy NGC 448}",
      journal = {\mnras},
         year = 2016,
        month = sep,
       volume = {461},
       number = {2},
        pages = {2068-2076},
          doi = {10.1093/mnras/stw1452},
archivePrefix = {arXiv},
       eprint = {1606.04862},
 primaryClass = {astro-ph.GA},
       adsurl = {https://ui.adsabs.harvard.edu/abs/2016MNRAS.461.2068K}
}

@ARTICLE{2016MNRAS.458.2371R,
       author = {{Rodriguez-Gomez}, Vicente and {Pillepich}, Annalisa and {Sales}, Laura V. and {Genel}, Shy and {Vogelsberger}, Mark and {Zhu}, Qirong and {Wellons}, Sarah and {Nelson}, Dylan and {Torrey}, Paul and {Springel}, Volker and {Ma}, Chung-Pei and {Hernquist}, Lars},
        title = "{The stellar mass assembly of galaxies in the Illustris simulation: growth by mergers and the spatial distribution of accreted stars}",
      journal = {\mnras},
         year = 2016,
        month = may,
       volume = {458},
       number = {3},
        pages = {2371-2390},
          doi = {10.1093/mnras/stw456},
archivePrefix = {arXiv},
       eprint = {1511.08804},
 primaryClass = {astro-ph.GA},
       adsurl = {https://ui.adsabs.harvard.edu/abs/2016MNRAS.458.2371R}
}

@ARTICLE{2016MNRAS.457.1419M,
       author = {{Monachesi}, Antonela and {Bell}, Eric F. and {Radburn-Smith}, David J. and {Bailin}, Jeremy and {de Jong}, Roelof S. and {Holwerda}, Benne and {Streich}, David and {Silverstein}, Grace},
        title = "{The GHOSTS survey - II. The diversity of halo colour and metallicity profiles of massive disc galaxies}",
      journal = {\mnras},
         year = 2016,
        month = apr,
       volume = {457},
       number = {2},
        pages = {1419-1446},
          doi = {10.1093/mnras/stv2987},
archivePrefix = {arXiv},
       eprint = {1507.06657},
 primaryClass = {astro-ph.GA},
       adsurl = {https://ui.adsabs.harvard.edu/abs/2016MNRAS.457.1419M}
}

@ARTICLE{2015A&C....13...12N,
       author = {{Nelson}, D. and {Pillepich}, A. and {Genel}, S. and {Vogelsberger}, M. and {Springel}, V. and {Torrey}, P. and {Rodriguez-Gomez}, V. and {Sijacki}, D. and {Snyder}, G.~F. and {Griffen}, B. and {Marinacci}, F. and {Blecha}, L. and {Sales}, L. and {Xu}, D. and {Hernquist}, L.},
        title = "{The illustris simulation: Public data release}",
      journal = {Astronomy and Computing},
         year = 2015,
        month = nov,
       volume = {13},
        pages = {12-37},
          doi = {10.1016/j.ascom.2015.09.003},
archivePrefix = {arXiv},
       eprint = {1504.00362},
 primaryClass = {astro-ph.CO},
       adsurl = {https://ui.adsabs.harvard.edu/abs/2015A&C....13...12N}
}

@ARTICLE{2015A&A...581A..65C,
       author = {{Coccato}, L. and {Fabricius}, M. and {Morelli}, L. and {Corsini}, E.~M. and {Pizzella}, A. and {Erwin}, P. and {Dalla Bont{\`a}}, E. and {Saglia}, R. and {Bender}, R. and {Williams}, M.},
        title = "{Properties and formation mechanism of the stellar counter-rotating components in NGC 4191}",
      journal = {\aap},
         year = 2015,
        month = sep,
       volume = {581},
          eid = {A65},
        pages = {A65},
          doi = {10.1051/0004-6361/201526560},
archivePrefix = {arXiv},
       eprint = {1507.07936},
 primaryClass = {astro-ph.GA},
       adsurl = {https://ui.adsabs.harvard.edu/abs/2015A&A...581A..65C}
}

@ARTICLE{2015MNRAS.451.3269V,
       author = {{van de Voort}, Freeke and {Davis}, Timothy A. and {Kere{\v{s}}}, Du{\v{s}}an and {Quataert}, Eliot and {Faucher-Gigu{\`e}re}, Claude-Andr{\'e} and {Hopkins}, Philip F.},
        title = "{The creation and persistence of a misaligned gas disc in a simulated early-type galaxy}",
      journal = {\mnras},
         year = 2015,
        month = aug,
       volume = {451},
       number = {3},
        pages = {3269-3277},
          doi = {10.1093/mnras/stv1217},
archivePrefix = {arXiv},
       eprint = {1504.03685},
 primaryClass = {astro-ph.GA},
       adsurl = {https://ui.adsabs.harvard.edu/abs/2015MNRAS.451.3269V}
}

@ARTICLE{2015MNRAS.449...49R,
       author = {{Rodriguez-Gomez}, Vicente and {Genel}, Shy and {Vogelsberger}, Mark and {Sijacki}, Debora and {Pillepich}, Annalisa and {Sales}, Laura V. and {Torrey}, Paul and {Snyder}, Greg and {Nelson}, Dylan and {Springel}, Volker and {Ma}, Chung-Pei and {Hernquist}, Lars},
        title = "{The merger rate of galaxies in the Illustris simulation: a comparison with observations and semi-empirical models}",
      journal = {\mnras},
         year = 2015,
        month = may,
       volume = {449},
       number = {1},
        pages = {49-64},
          doi = {10.1093/mnras/stv264},
archivePrefix = {arXiv},
       eprint = {1502.01339},
 primaryClass = {astro-ph.GA},
       adsurl = {https://ui.adsabs.harvard.edu/abs/2015MNRAS.449...49R}
}

@ARTICLE{2015ApJ...804L..40G,
       author = {{Genel}, Shy and {Fall}, S. Michael and {Hernquist}, Lars and {Vogelsberger}, Mark and {Snyder}, Gregory F. and {Rodriguez-Gomez}, Vicente and {Sijacki}, Debora and {Springel}, Volker},
        title = "{Galactic Angular Momentum in the Illustris Simulation: Feedback and the Hubble Sequence}",
      journal = {\apjl},
         year = 2015,
        month = may,
       volume = {804},
       number = {2},
          eid = {L40},
        pages = {L40},
          doi = {10.1088/2041-8205/804/2/L40},
archivePrefix = {arXiv},
       eprint = {1503.01117},
 primaryClass = {astro-ph.GA},
       adsurl = {https://ui.adsabs.harvard.edu/abs/2015ApJ...804L..40G}
}

@ARTICLE{2014A&A...570A..79P,
       author = {{Pizzella}, A. and {Morelli}, L. and {Corsini}, E.~M. and {Dalla Bont{\`a}}, E. and {Coccato}, L. and {Sanjana}, G.},
        title = "{The difference in age of the two counter-rotating stellar disks of the spiral galaxy NGC 4138}",
      journal = {\aap},
         year = 2014,
        month = oct,
       volume = {570},
          eid = {A79},
        pages = {A79},
          doi = {10.1051/0004-6361/201424746},
archivePrefix = {arXiv},
       eprint = {1409.3086},
 primaryClass = {astro-ph.GA},
       adsurl = {https://ui.adsabs.harvard.edu/abs/2014A&A...570A..79P}
}

@ARTICLE{2014MNRAS.437.3596A,
       author = {{Algorry}, David G. and {Navarro}, Julio F. and {Abadi}, Mario G. and {Sales}, Laura V. and {Steinmetz}, Matthias and {Piontek}, Franziska},
        title = "{Counterrotating stars in simulated galaxy discs}",
      journal = {\mnras},
         year = 2014,
        month = feb,
       volume = {437},
       number = {4},
        pages = {3596-3602},
          doi = {10.1093/mnras/stt2154},
archivePrefix = {arXiv},
       eprint = {1311.1215},
 primaryClass = {astro-ph.CO},
       adsurl = {https://ui.adsabs.harvard.edu/abs/2014MNRAS.437.3596A}
}

@ARTICLE{2013ApJ...769..105K,
       author = {{Katkov}, Ivan Yu. and {Sil'chenko}, Olga K. and {Afanasiev}, Victor L.},
        title = "{Lenticular Galaxy IC 719: Current Building of the Counterrotating Large-scale Stellar Disk}",
      journal = {\apj},
         year = 2013,
        month = jun,
       volume = {769},
       number = {2},
          eid = {105},
        pages = {105},
          doi = {10.1088/0004-637X/769/2/105},
archivePrefix = {arXiv},
       eprint = {1304.3339},
 primaryClass = {astro-ph.CO},
       adsurl = {https://ui.adsabs.harvard.edu/abs/2013ApJ...769..105K}
}

@ARTICLE{2013A&A...549A...3C,
       author = {{Coccato}, L. and {Morelli}, L. and {Pizzella}, A. and {Corsini}, E.~M. and {Buson}, L.~M. and {Dalla Bont{\`a}}, E.},
        title = "{Spectroscopic evidence of distinct stellar populations in the counter-rotating stellar disks of NGC 3593 and NGC 4550}",
      journal = {\aap},
         year = 2013,
        month = jan,
       volume = {549},
          eid = {A3},
        pages = {A3},
          doi = {10.1051/0004-6361/201220460},
archivePrefix = {arXiv},
       eprint = {1210.7807},
 primaryClass = {astro-ph.CO},
       adsurl = {https://ui.adsabs.harvard.edu/abs/2013A&A...549A...3C}
}

@ARTICLE{2013MNRAS.428.1296J,
       author = {{Johnston}, Evelyn J. and {Merrifield}, Michael R. and {Arag{\'o}n-Salamanca}, Alfonso and {Cappellari}, Michele},
        title = "{Disentangling the stellar populations in the counter-rotating disc galaxy NGC 4550}",
      journal = {\mnras},
         year = 2013,
        month = jan,
       volume = {428},
       number = {2},
        pages = {1296-1302},
          doi = {10.1093/mnras/sts121},
archivePrefix = {arXiv},
       eprint = {1210.0535},
 primaryClass = {astro-ph.CO},
       adsurl = {https://ui.adsabs.harvard.edu/abs/2013MNRAS.428.1296J}
}

@ARTICLE{2012MNRAS.420..255T,
       author = {{Tissera}, Patricia B. and {White}, Simon D.~M. and {Scannapieco}, Cecilia},
        title = "{Chemical signatures of formation processes in the stellar populations of simulated galaxies}",
      journal = {\mnras},
         year = 2012,
        month = feb,
       volume = {420},
       number = {1},
        pages = {255-270},
          doi = {10.1111/j.1365-2966.2011.20028.x},
archivePrefix = {arXiv},
       eprint = {1110.5864},
 primaryClass = {astro-ph.CO},
       adsurl = {https://ui.adsabs.harvard.edu/abs/2012MNRAS.420..255T}
}

@ARTICLE{2011MNRAS.412L.113C,
       author = {{Coccato}, L. and {Morelli}, L. and {Corsini}, E.~M. and {Buson}, L. and {Pizzella}, A. and {Vergani}, D. and {Bertola}, F.},
        title = "{Dating the formation of the counter-rotating stellar disc in the spiral galaxy NGC 5719 by disentangling its stellar populations}",
      journal = {\mnras},
         year = 2011,
        month = mar,
       volume = {412},
       number = {1},
        pages = {L113-L117},
          doi = {10.1111/j.1745-3933.2011.01016.x},
archivePrefix = {arXiv},
       eprint = {1101.3092},
 primaryClass = {astro-ph.GA},
       adsurl = {https://ui.adsabs.harvard.edu/abs/2011MNRAS.412L.113C}
}

@ARTICLE{2010MNRAS.408..783R,
       author = {{Ro{\v{s}}kar}, Rok and {Debattista}, Victor P. and {Brooks}, Alyson M. and {Quinn}, Thomas R. and {Brook}, Chris B. and {Governato}, Fabio and {Dalcanton}, Julianne J. and {Wadsley}, James},
        title = "{Misaligned angular momentum in hydrodynamic cosmological simulations: warps, outer discs and thick discs}",
      journal = {\mnras},
         year = 2010,
        month = oct,
       volume = {408},
       number = {2},
        pages = {783-796},
          doi = {10.1111/j.1365-2966.2010.17178.x},
archivePrefix = {arXiv},
       eprint = {1006.1659},
 primaryClass = {astro-ph.CO},
       adsurl = {https://ui.adsabs.harvard.edu/abs/2010MNRAS.408..783R}
}

@ARTICLE{2010ApJ...710L.156R,
       author = {{Rupke}, David S.~N. and {Kewley}, Lisa J. and {Barnes}, Joshua E.},
        title = "{Galaxy Mergers and the Mass-Metallicity Relation: Evidence for Nuclear Metal Dilution and Flattened Gradients from Numerical Simulations}",
      journal = {\apjl},
         year = 2010,
        month = feb,
       volume = {710},
       number = {2},
        pages = {L156-L160},
          doi = {10.1088/2041-8205/710/2/L156},
archivePrefix = {arXiv},
       eprint = {1001.1728},
 primaryClass = {astro-ph.GA},
       adsurl = {https://ui.adsabs.harvard.edu/abs/2010ApJ...710L.156R}
}

@ARTICLE{2010MNRAS.401..791S,
       author = {{Springel}, Volker},
        title = "{E pur si muove: Galilean-invariant cosmological hydrodynamical simulations on a moving mesh}",
      journal = {\mnras},
         year = 2010,
        month = jan,
       volume = {401},
       number = {2},
        pages = {791-851},
          doi = {10.1111/j.1365-2966.2009.15715.x},
archivePrefix = {arXiv},
       eprint = {0901.4107},
 primaryClass = {astro-ph.CO},
       adsurl = {https://ui.adsabs.harvard.edu/abs/2010MNRAS.401..791S}
}

@ARTICLE{2009MNRAS.393.1255C,
       author = {{Crocker}, Alison F. and {Jeong}, Hyunjin and {Komugi}, Shinya and {Combes}, Francoise and {Bureau}, Martin and {Young}, Lisa M. and {Yi}, Sukyoung},
        title = "{Molecular gas and star formation in the red-sequence counter-rotating disc galaxy NGC 4550}",
      journal = {\mnras},
         year = 2009,
        month = mar,
       volume = {393},
       number = {4},
        pages = {1255-1264},
          doi = {10.1111/j.1365-2966.2008.14295.x},
archivePrefix = {arXiv},
       eprint = {0812.0178},
 primaryClass = {astro-ph},
       adsurl = {https://ui.adsabs.harvard.edu/abs/2009MNRAS.393.1255C}
}

@ARTICLE{2007A&A...463..883V,
       author = {{Vergani}, D. and {Pizzella}, A. and {Corsini}, E.~M. and {van Driel}, W. and {Buson}, L.~M. and {Dettmar}, R.-J. and {Bertola}, F.},
        title = "{NGC 5719/13: interacting spirals forming a counter-rotating stellar disc}",
      journal = {\aap},
         year = 2007,
        month = mar,
       volume = {463},
       number = {3},
        pages = {883-892},
          doi = {10.1051/0004-6361:20066413},
archivePrefix = {arXiv},
       eprint = {astro-ph/0611426},
 primaryClass = {astro-ph},
       adsurl = {https://ui.adsabs.harvard.edu/abs/2007A&A...463..883V}
}

@ARTICLE{2006PASP..118..517B,
       author = {{Boselli}, Alessandro and {Gavazzi}, Giuseppe},
        title = "{Environmental Effects on Late-Type Galaxies in Nearby Clusters}",
      journal = {\pasp},
         year = 2006,
        month = apr,
       volume = {118},
       number = {842},
        pages = {517-559},
          doi = {10.1086/500691},
archivePrefix = {arXiv},
       eprint = {astro-ph/0601108},
 primaryClass = {astro-ph},
       adsurl = {https://ui.adsabs.harvard.edu/abs/2006PASP..118..517B}
}

@ARTICLE{2004A&A...424..447P,
       author = {{Pizzella}, A. and {Corsini}, E.~M. and {Vega Beltr{\'a}n}, J.~C. and {Bertola}, F.},
        title = "{Ionized gas and stellar kinematics of seventeen nearby spiral galaxies}",
      journal = {\aap},
         year = 2004,
        month = sep,
       volume = {424},
        pages = {447-454},
          doi = {10.1051/0004-6361:20047183},
archivePrefix = {arXiv},
       eprint = {astro-ph/0404558},
 primaryClass = {astro-ph},
       adsurl = {https://ui.adsabs.harvard.edu/abs/2004A&A...424..447P}
}

@ARTICLE{2003ApJ...597...21A,
       author = {{Abadi}, Mario G. and {Navarro}, Julio F. and {Steinmetz}, Matthias and {Eke}, Vincent R.},
        title = "{Simulations of Galaxy Formation in a {\ensuremath{\Lambda}} Cold Dark Matter Universe. II. The Fine Structure of Simulated Galactic Disks}",
      journal = {\apj},
         year = 2003,
        month = nov,
       volume = {597},
       number = {1},
        pages = {21-34},
          doi = {10.1086/378316},
archivePrefix = {arXiv},
       eprint = {astro-ph/0212282},
 primaryClass = {astro-ph},
       adsurl = {https://ui.adsabs.harvard.edu/abs/2003ApJ...597...21A}
}

@ARTICLE{2002MNRAS.333..327T,
       author = {{Tissera}, P.~B. and {Dom{\'\i}nguez-Tenreiro}, R. and {Scannapieco}, C. and {S{\'a}iz}, A.},
        title = "{Double starbursts triggered by mergers in hierarchical clustering scenarios}",
      journal = {\mnras},
         year = 2002,
        month = jun,
       volume = {333},
       number = {2},
        pages = {327-338},
          doi = {10.1046/j.1365-8711.2002.05385.x},
archivePrefix = {arXiv},
       eprint = {astro-ph/0202160},
 primaryClass = {astro-ph},
       adsurl = {https://ui.adsabs.harvard.edu/abs/2002MNRAS.333..327T}
}

@ARTICLE{2001Ap&SS.276..909P,
       author = {{Puerari}, I. and {Pfenniger}, D.},
        title = "{Formation of Massive Counter-Rotating Discs: An Alternative Scenario}",
      journal = {\apss},
         year = 2001,
        month = mar,
       volume = {276},
        pages = {909-914},
          doi = {10.1023/A:1017581325673},
archivePrefix = {arXiv},
       eprint = {astro-ph/9903096},
 primaryClass = {astro-ph},
       adsurl = {https://ui.adsabs.harvard.edu/abs/2001Ap&SS.276..909P}
}

@ARTICLE{2000A&A...363..869G,
       author = {{Garc{\'\i}a-Burillo}, S. and {Sempere}, M.~J. and {Combes}, F. and {Hunt}, L.~K. and {Neri}, R.},
        title = "{Anatomy of the counterrotating molecular disk in the spiral NGC 3593. $^{12}$CO(1-0) interferometer observations and numerical simulations}",
      journal = {\aap},
         year = 2000,
        month = nov,
       volume = {363},
        pages = {869-886},
       adsurl = {https://ui.adsabs.harvard.edu/abs/2000A&A...363..869G}
}

@ARTICLE{1998A&A...337...80C,
       author = {{Corsini}, E.~M. and {Pizzella}, A. and {Funes}, J.~G. and {Vega Beltran}, J.~C. and {Bertola}, F.},
        title = "{The circumnuclear ring of ionized gas in NGC 3593}",
      journal = {\aap},
         year = 1998,
        month = sep,
       volume = {337},
        pages = {80-84},
          doi = {10.48550/arXiv.astro-ph/9806066},
archivePrefix = {arXiv},
       eprint = {astro-ph/9806066},
 primaryClass = {astro-ph},
       adsurl = {https://ui.adsabs.harvard.edu/abs/1998A&A...337...80C}
}

@INPROCEEDINGS{1996AAS...189.6804J,
       author = {{Jore}, K.~P. and {Haynes}, M.~P. and {Broeils}, A.~H.},
        title = "{The HI Distribution and Kinematics in SA Galaxies}",
    booktitle = {American Astronomical Society Meeting Abstracts},
         year = 1996,
       series = {American Astronomical Society Meeting Abstracts},
       volume = {189},
        month = dec,
          eid = {68.04},
        pages = {68.04},
       adsurl = {https://ui.adsabs.harvard.edu/abs/1996AAS...189.6804J}
}

@ARTICLE{1996MNRAS.283..543K,
       author = {{Kuijken}, K. and {Fisher}, D. and {Merrifield}, M.~R.},
        title = "{A search for counter-rotating stars in S0 galaxies.}",
      journal = {\mnras},
         year = 1996,
        month = dec,
       volume = {283},
       number = {2},
        pages = {543-550},
          doi = {10.1093/mnras/283.2.543},
archivePrefix = {arXiv},
       eprint = {astro-ph/9606099},
 primaryClass = {astro-ph},
       adsurl = {https://ui.adsabs.harvard.edu/abs/1996MNRAS.283..543K}
}

@ARTICLE{1996ApJ...471..115B,
       author = {{Barnes}, Joshua E. and {Hernquist}, Lars},
        title = "{Transformations of Galaxies. II. Gasdynamics in Merging Disk Galaxies}",
      journal = {\apj},
         year = 1996,
        month = nov,
       volume = {471},
        pages = {115},
          doi = {10.1086/177957},
       adsurl = {https://ui.adsabs.harvard.edu/abs/1996ApJ...471..115B}
}

@ARTICLE{1996ApJ...461...55T,
       author = {{Thakar}, Aniruddha R. and {Ryden}, Barbara S.},
        title = "{Formation of Massive Counterrotating Disks in Spiral Galaxies}",
      journal = {\apj},
         year = 1996,
        month = apr,
       volume = {461},
        pages = {55},
          doi = {10.1086/177037},
archivePrefix = {arXiv},
       eprint = {astro-ph/9510053},
 primaryClass = {astro-ph},
       adsurl = {https://ui.adsabs.harvard.edu/abs/1996ApJ...461...55T}
}

@ARTICLE{1996ApJ...458L..67B,
       author = {{Bertola}, Francesco and {Cinzano}, Pierantonio and {Corsini}, Enrico Maria and {Pizzella}, Alessandro and {Persic}, Massimo and {Salucci}, Paolo},
        title = "{Counterrotating Stellar Disks in Early-Type Spirals: NGC 3593}",
      journal = {\apjl},
         year = 1996,
        month = feb,
       volume = {458},
        pages = {L67},
          doi = {10.1086/309924},
       adsurl = {https://ui.adsabs.harvard.edu/abs/1996ApJ...458L..67B}
}

@ARTICLE{1994ApJ...432..575M,
       author = {{Merrifield}, Michael R. and {Kuijken}, Konrad},
        title = "{Counterrotating Stars in the Disk of the SAB Galaxy NGC 7217}",
      journal = {\apj},
         year = 1994,
        month = sep,
       volume = {432},
        pages = {575},
          doi = {10.1086/174596},
       adsurl = {https://ui.adsabs.harvard.edu/abs/1994ApJ...432..575M}
}

@ARTICLE{1994AJ....108..456R,
       author = {{Rubin}, Vera C.},
        title = "{Multi-Spin Galaxies}",
      journal = {\aj},
         year = 1994,
        month = aug,
       volume = {108},
        pages = {456},
          doi = {10.1086/117083},
       adsurl = {https://ui.adsabs.harvard.edu/abs/1994AJ....108..456R}
}

@ARTICLE{1994ApJ...420..558B,
       author = {{Braun}, Robert and {Walterbos}, Rene A.~M. and {Kennicutt}, Jr., Robert C. and {Tacconi}, Linda J.},
        title = "{Counterrotating Gaseous Disks in NGC 4826}",
      journal = {\apj},
         year = 1994,
        month = jan,
       volume = {420},
        pages = {558},
          doi = {10.1086/173586},
       adsurl = {https://ui.adsabs.harvard.edu/abs/1994ApJ...420..558B}
}

@ARTICLE{1992ApJ...400L...5R,
       author = {{Rix}, Hans-Walter and {Franx}, Marijn and {Fisher}, David and {Illingworth}, Garth},
        title = "{NGC 4550: A Laboratory for Testing Galaxy Formation}",
      journal = {\apjl},
         year = 1992,
        month = nov,
       volume = {400},
        pages = {L5},
          doi = {10.1086/186635},
       adsurl = {https://ui.adsabs.harvard.edu/abs/1992ApJ...400L...5R}
}

@ARTICLE{1992ApJ...394L...9R,
       author = {{Rubin}, Vera C. and {Graham}, J.~A. and {Kenney}, Jeffrey D.~P.},
        title = "{Cospatial Counterrotating Stellar Disks in the Virgo E7/S0 Galaxy NGC 4550}",
      journal = {\apjl},
         year = 1992,
        month = jul,
       volume = {394},
        pages = {L9},
          doi = {10.1086/186460},
       adsurl = {https://ui.adsabs.harvard.edu/abs/1992ApJ...394L...9R}
}

\begin{appendix}
    \onecolumn
    \section{Spatial distribution of components for galaxies hosting a significant CRD}
    \label{sec:appendix_spatialdist}
\begin{figure*}[h]
            \centering
            \includegraphics[width=6.8in]{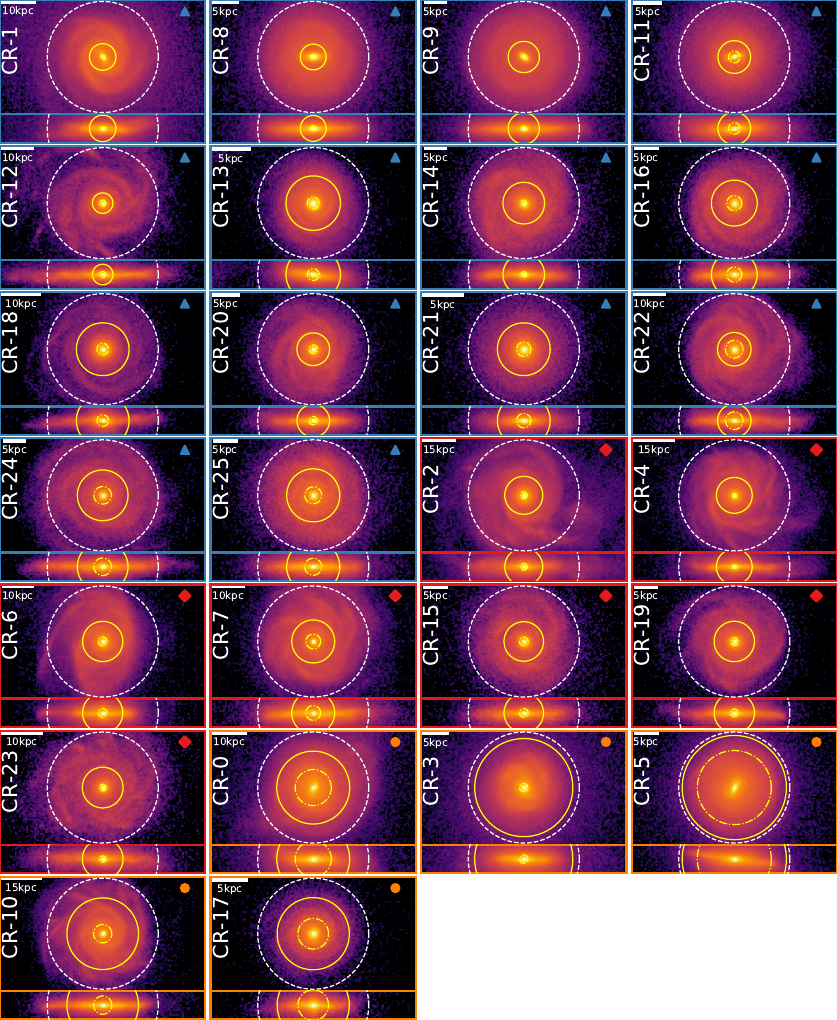}
            \caption{Face-on and edge-on projections of the stellar mass surface density for the 26 significant CRD galaxies. The galaxies are sorted as in figure \ref{fig:ages_CRDs}. We show all the stellar particles that belong to the host galaxy. The solid and dotted-dashed yellow circles represent the radii, $R_{95}^{\rm CRD}$ and $R_{50}^{\rm CRD}$, of the coplanar counterrotating disk, respectively, while the white dashed circle represents the optical radius, $R_{\rm opt}$, of the galaxy. In the top-right corner of each panel, the type is shown, which are extended ex situ CRDs, compact ex situ CRDs, compact in situ CRDs, extended in situ CRDs. The types are respectively represented with a yellow star, red rhombuses, blue triangles, and orange circles.}
            \label{fig:CRDs_all_appendix}
        \end{figure*}

\begin{figure*}[ht]
            \centering
            \includegraphics[width=6.8in]{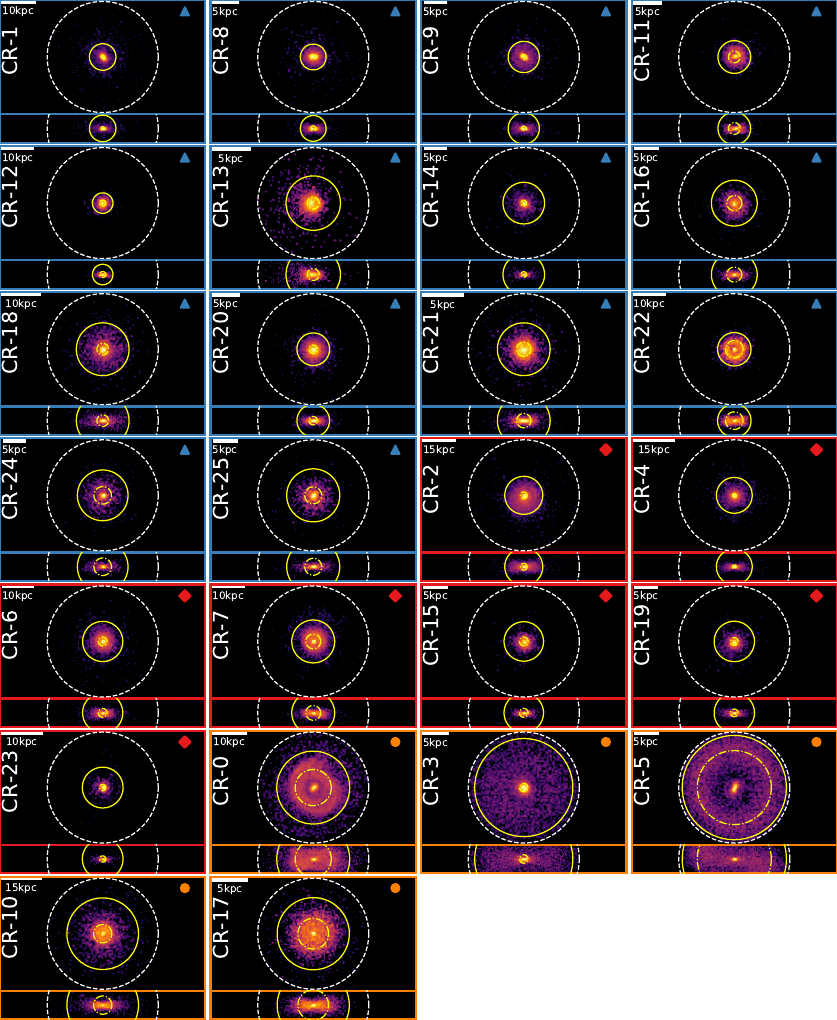}
            \caption{Same as in \ref{fig:CRDs_all_appendix} but for only counterrotating stellar disk particles, as described in Section \ref{section:crdef}}
            \label{fig:CRDs_CRD_appendix}
        \end{figure*}

        Figures \ref{fig:CRDs_all_appendix} and \ref{fig:CRDs_CRD_appendix} show the face-on and edge-on projections of the stellar mass surface density for the 26 significant CRD galaxies. We show the scale size bar and the internal id on the top-left corner and left margin of each panel, respectively. The variety of configurations for the CRDs can also be appreciated.
        
        \newpage
    \twocolumn
    \section{Orbital properties of infalling satellites in compact in situ CRDs.}
    \label{sec:appendix_extra_infalling_satellites}
\begin{figure}[ht]
            \centering
            \includegraphics[width=3.2in]{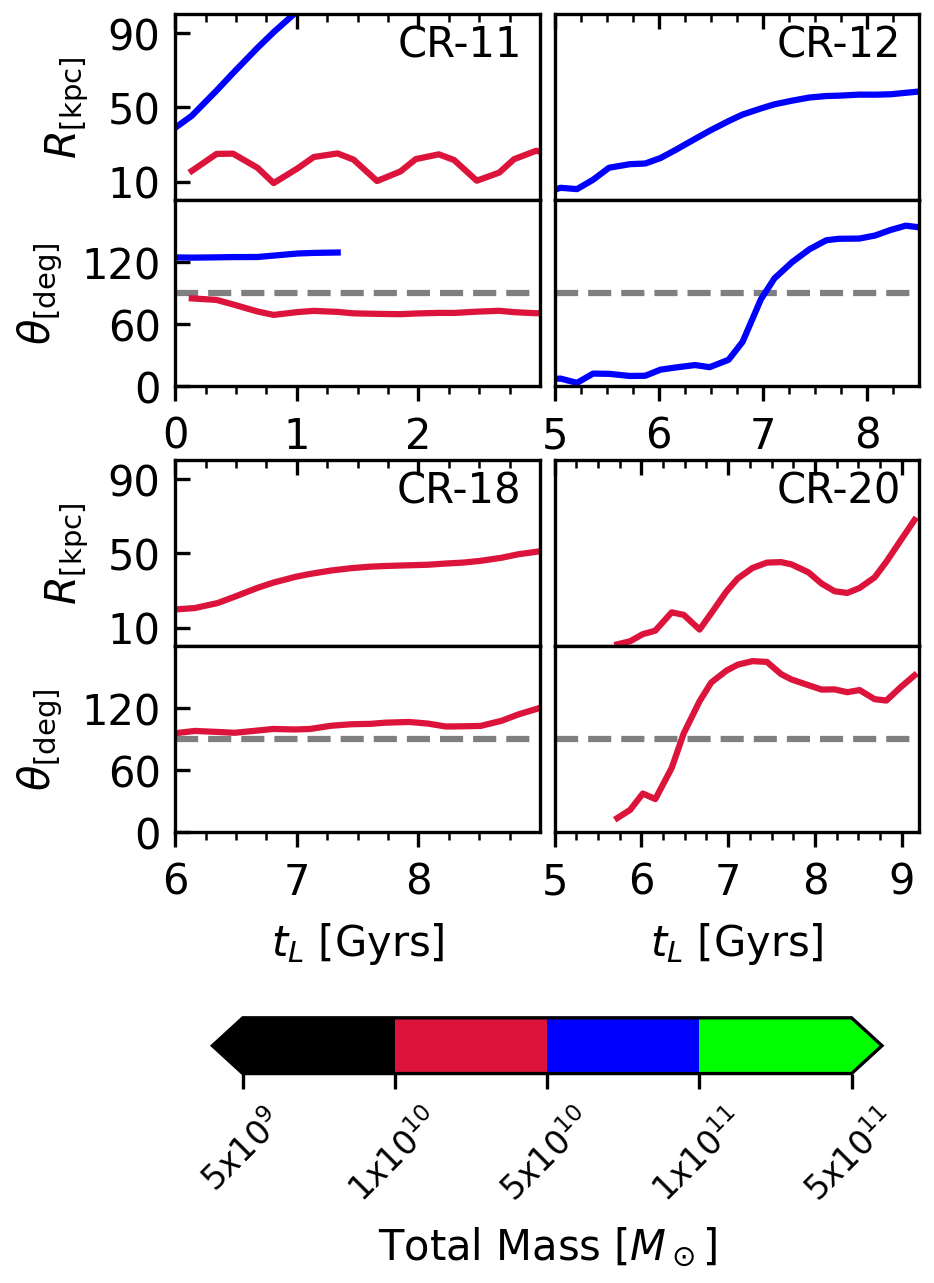}
            \caption{The orbit of the satellites that are in interaction with CR-11, CR-12, CR-18 and CR-20, respectively. Each column of panels is constrained to the lookback time window that has a peak in the SFH fo the CRD according to Fig. \ref{fig:ages_CRDs}. Top panel: evolution of the satellite galactocentric distance from the host. Bottom panel: evolution of the angle between the angular momentum of the disk and the satellite's orbital angular momentum. Orbits near $180^\circ$ are counterrotating, while those near $0^\circ$ are corotating.}
            \label{fig:sat_info_appendix}
        \end{figure}

        Figure \ref{fig:sat_info_appendix} shows the orbital properties for compact in situ CRDs that formed later, with their counterrotating stars born during one or more bursts triggered by satellite interactions (but without significant stellar accretion). The colors represent each orbit of a satellite and are the same as in Fig. \ref{fig:ages_CRDs}. It can be appreciated how in each case at least one satellite initially had a misaligned orbit.

        \newpage
    \section{Differences with past studies on counterrotating disks}
    \label{sec:appendix_differences_with_past_studies}
\begin{figure}[ht]
            \centering
            \includegraphics[width=3.2in]{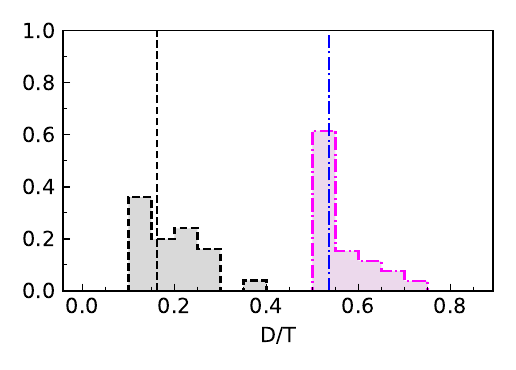}
            \caption{Distribution of D/T values of galaxy models with identified CRDs, extracted from the publicly available catalogs from the IllustrisTNG project \cite{2015ApJ...804L..40G}. The gray (dashed) and magenta (dash-dotted) histograms indicate the D/T distribution of CRDs galaxies identified in \citealt{2021MNRAS.500.3870K} and Bugueño et al. (2025), respectively.}
            \label{fig:DT_comparison_khop_bugueno}
        \end{figure}

    As mentioned in Sec. \ref{section:intro} of this paper, there have been a multitude of studies on the topic of counterrotating disks. Most of them, however, have focused on lenticular and elliptical galaxies where these types of structures are more common. Here, we compare a study by \cite{2021MNRAS.500.3870K} with our work. In Figure \ref{fig:DT_comparison_khop_bugueno}, we show the distribution of the disk-to-total mass ratio (D/T) of the counterrotating galaxy sample selected in \cite{2021MNRAS.500.3870K}, according to \cite{2015ApJ...804L..40G}'s catalog, and the sample in this work.

    We observed that the galaxies selected in this work are characterized by a higher D/T, meaning that they include late-type disk galaxies which are different from the early-type disk counterparts studies in \cite{2021MNRAS.500.3870K}. This also means that caution is required when comparing our results with observational studies, which have mainly studied CRDs in lenticular galaxies (\citealt{2022MNRAS.511..139B}; \citealt{2025ApJS..281...19G}; \citealt{2013ApJ...769..105K}; \citealt{2018MNRAS.475..648P}; \citealt{2013A&A...549A...3C}; \citealt{2013MNRAS.428.1296J}; \citealt{2015A&A...581A..65C}; \citealt{2016MNRAS.461.2068K}; \citealt{2017A&A...600A..76M}; \citealt{2017MNRAS.464.4789M}).

    We note that, in this work, we also adopted a definition of CRDs based on the circularity parameter, similarly to \cite{2003ApJ...597...21A}, which is more restrictive with respect to the definition of CRDs adopted by \citealt{2021MNRAS.500.3870K} based on the circular velocity. For this reason, in this work, we find that CRDs in late-type galaxies generally comprise a small fraction of the stellar mass with respect to the whole disk, which is more consistent with the results of \citealt{2022MNRAS.509.1764J} based on the TNG100 simulations.
\end{appendix}

\end{document}